# Using the Parameterized Quantum Circuit combined with Variational-Quantum-Eigensolver (VQE) to create an Intelligent social workers' schedule problem solver

Atchade Parfait Adelomou [1] Elisabet Golobardes Ribé [2] and Xavier Vilasis Cardona[3]

[1,2,3] Research Group in Data Science for the Digital Society (DS4DS) La Salle – U. Ramon Llull, Sant Joan de la Salle, 42, 08022 Barcelona, Spain

parfait.atchade@salle.url.edu +34 661205543
https://orcid.org/0000-0003-2725-277X

**Abstract.** The social worker scheduling problem is a class of combinatorial optimization problems that combines scheduling with routing issues. These types of problems with classical computing can only be solved, in the best of cases, in an approximate way and significantly when the input data does not grow considerably. Today, the focus on the quantum computer should no longer be only on its enormous computing power, but also on the use of its imperfection for this era, (Noisy Intermediate-Scale Quantum (NISQ)) to create a powerful optimization and learning device that uses variational techniques. We had already proposed a formulation and solution of this problem using the capacity of the quantum computer. In this article, we present some broad results of the experimentation techniques. And above all, we propose an adaptive and intelligence solution, which efficiently recalculates the schedules of social workers. Taking into account new restrictions and changes in the initial conditions, by using a case-based reasoning system and the variational quantum eigensolver based on a finite-depth quantum circuit. That encodes the ground state of the Hamiltonian of social workers.

The quantum feasibility of the algorithm will be modelled with docplex and tested on IBMQ computers.

## 1 Introduction

The problem of social workers is given by a definite difficulty in generating optimal visiting hours. With the main objective to achieve optimal hours, each social worker can visit patients at home, provide them with personalized attention and assistance according to a schedule, time and duration of the visit depending on the patient's pathology. The social worker task combines, on the one hand, a routing problem such as, for example, a Vehicle Routing Problem (VRP) [1] [2] [3] and, on the other hand, a problem of planning [4]. Where the exact solution to this task can lead to exponential computation time when scaling the input data. Therefore, a different approach from classical computation is truly crucial to resolution. The complexity and especially the importance of this dilemma, involved the scientific community in the investigation of efficient methods to solve them [5].
New restrictions condition the problem of social workers' issue that was formulated by us [6]. Until now, the optimal schedule was calculated every day or every moment when the end-user required it. In this way, any changes suffered by the patient, the social worker or the administration system (town halls, hospitals, nursing homes, etc.) could be introduced.
In this case, the town hall is considered as a management system.
The latter decides that, given the emergency cases of the new digital therapies [7], patients will not be able to change social workers daily, but every three months or when the degree of degradation of optimal schedules is alarming. The changes in registrations and cancellations of both patients and social workers will be modified daily. However, they must wait for three months established to enter the recalculation of the search for the best optimal schedule for social workers. Facing this new reality, it has been decided to refocus the solution to the problem without redefining the formulation of social workers from scratch.
To do this, we propose a new approach that guarantees, on the one hand, using the necessary and fair computational resources efficiently by using the previous data stored together with the new inputs to solve the problem without having to recalculate everything from the beginning (Top-Down philosophy). And, on the other hand, propose a hybrid basis (quantum-classical) to generalize the resolution of routing and planning problems based on cases [8] [9], i.e. using a case-based reasoning philosophy [10].

The article is organized as follows. In Section 2, the importance of the ansatz - Parameterized Quantum Circuits (PQC) [11], its characteristics, and its choice for the design of a variational system will be explored. In section 3, the Variational Quantum Eigensolver VQE, which is the algorithm used to find the expected values of the Hamiltonian that maps the objective function will be introduced. In Section 4, how PQC can be the basis of Machine Learning (ML) will be analyzed. In Section 5, the question to solve and its approaches will be presented. In sections 6, 7 and 8, the formulation of the social workers proposed to be solved in a quantum computer will be recovered. Finally, in Section 9, the results of the experimentation will be shared. In the discussion, in addition to commenting on the results obtained, let us discuss based on comparative results, how the approach proposed in this article can be a solid basis for generalizing the resolution of specific problems. Taking into account that these problems may undergo some changes in requirements by the city council without changing the essence of the initial project.

## 2 The ansatz - Parameterized quantum circuits (PQC)

The basic idea that one pursues is to have an ansatz that, formed by basic gates for quantum computing, is the most representative in the Hilbert vector space. In other words, with the control or parameterization of these parameters, the ansatz, in particular, maps the maximum number of points within the Bloch sphere (Fig. 1). Another way for looking at it is understanding the objective of the ansatz is to find the state vector that best approximates all the points of the Hilbert vector space. We just have to remember that the Hilbert vector space is the computational space of quantum mechanics, therefore, of quantum computing.

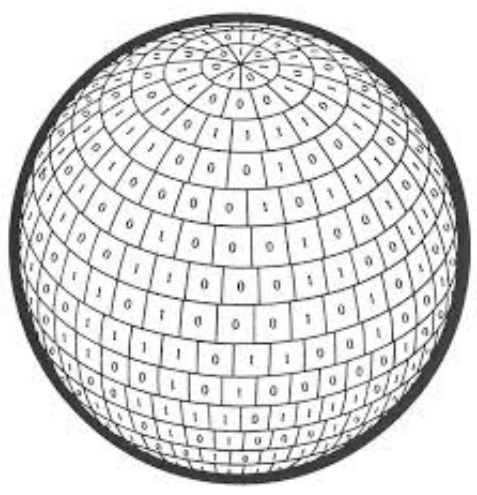

**Fig. 1.** Bloch sphere, where the infinite Hilbert vector space resides. It is worth remembering that Hilbert space is the immense place where the states that describe a quantum system live. Let $N$ be the number of qubits and let $2^N$ be the space dimension. Given the 65-qubit IBM processor, it will be $3{,}68935 \times 10^{19}$ dimensions. Many literatures lead us to use or, at least to think about using the quantum computer in Machine Learning (ML). Simply, by definition, the Hilbert vector space has defined these operations (internal and external products and mapping inputs in a large space) that a quantum computer performs natively and very easily [12].

Ansatz is based on a layered gate architecture, where the layer (known as a "block") is a sequence of gates that is repeated (Fig. 2). And, the number of said repetitions, forms a hyperparameter of the variational circuit at the same time that they represent its depth. The balance between the quantum parameterized circuit and its depth is key to the resolution and accuracy of variational algorithms.

Many literatures have reviewed the design and use of parameterized circuits to solve quantum problems. It is true that, by the nature of the problem, one circuit may be more useful than another, but it is not always clear what makes an ansatz suitable. The figure 3, presents the pattern in the construction of PQC where, given two unitaries $A(\alpha)$ and $B(\beta)$, $A(\alpha)$ is parameterized and $B(\beta)$ is fixed.

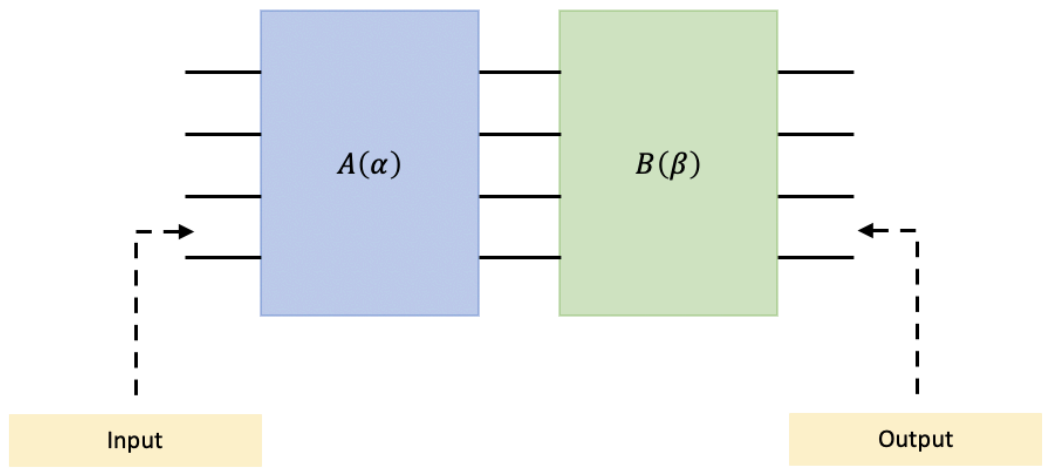

**Fig. 2.** A conceptual diagram that describes the pattern in the construction of quantum parameterized circuits. Where $A(\alpha)$ and $B(\beta)$ are unitaries

The basic intuition would be to achieve an ansatz with the quantum gates with the maximum number of controls (parameters). Since, with this, one would find the best approximate vector within the Hilbert vector space. To do this, on the one hand, Trotterization techniques [13] must be met at the layer level and, on the other, that the quantum gates comply with the Solovay Kitaev algorithm [14], which leads us to the fact that this intuition is not so trivial. The Trotterized technique was used in [15] for ADAPT-VQE to determine a quasi-optimal ansatz with the minimum number of operators for a desired level of precision.
There are several interesting templates developed under the Pennylane to create ansatz for some given problems [16]. Today, the best quantum gate for computing is $U_3$ (3 control parameters $\theta, \emptyset, \lambda$) within the Lie group $SU(2)$. The counterpart of this strategy is the number of control variables to manage.

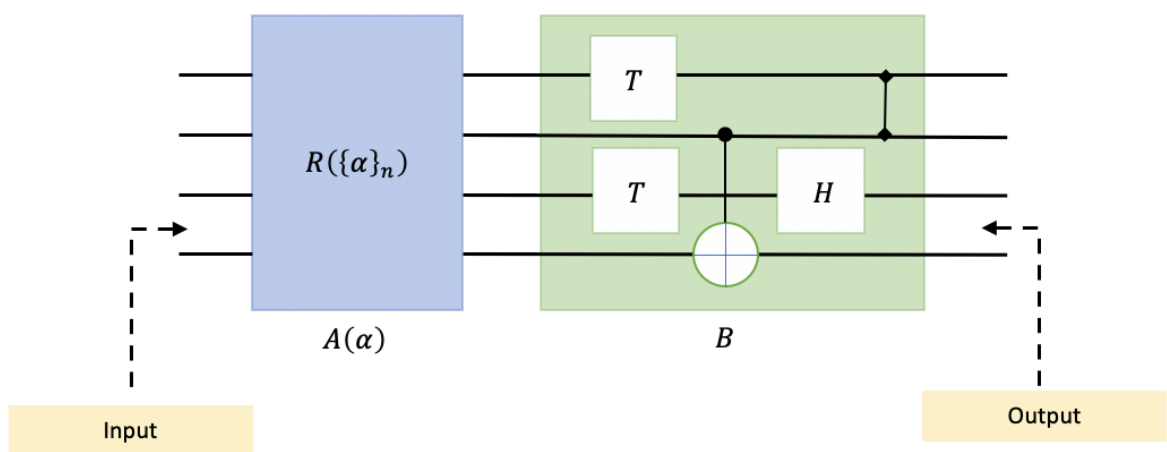

**Fig. 3.** A conceptual diagram that describes the pattern in the construction of quantum parameterized circuits. Where $A(\alpha)$ is parameterized and $B$ is fixed. $A(\alpha)$ and $B$ are unitaries

Within the proposed strategy for creating parameterized circuits, we will focus on fixed single-qubit gates. It is the circuit called Instantaneous Quantum Polynomial (IQP) [17], where $A$ consists of Hadamard gates and $B$ is formed by parameterized diagonal gates of one and two qubits (Fig. 4).

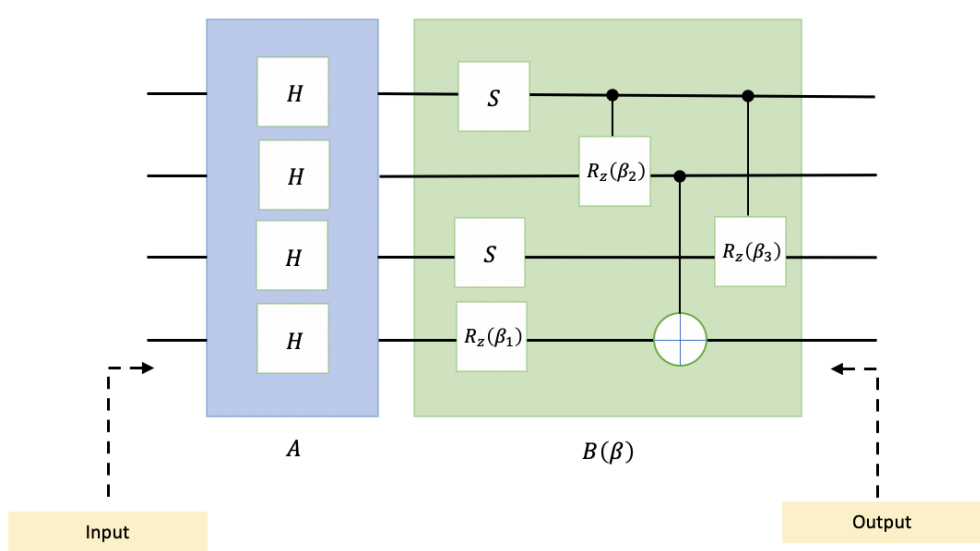

**Fig. 4.** A conceptual diagram that describes the pattern in the construction of quantum parameterized circuits. Both $A$ and $B(\beta)$ are two unitary. This circuit is known as an instantaneous quantum polynomial (IQP), where A consists of Hadamard gates and $B(\beta)$ is formed by parameterized diagonal gates of one and two qubits.

When deciding whether to recalculate a social worker schedule, its viability is checked, or a similar case is retrieved (Fig. 11.). With this model, we can approximately generate a state vector $|\psi\rangle$ in the Hilbert vector space that helps to calculate the minimum eigenvalue of the Hamiltonian associated with the problem of social workers. This minimum eigenvalue represents the ground state energy of the social workers' Hamiltonian.
However, it is emphasized that one of the flagship algorithms for architects based on gates models (for example, IBM, Microsoft, Alibaba, Google, Rigetti and Xanadu), Quantum Approximate Optimization Algorithm (QAOA) [18], is based on the variational circuit technique as we can see in the figure (5).

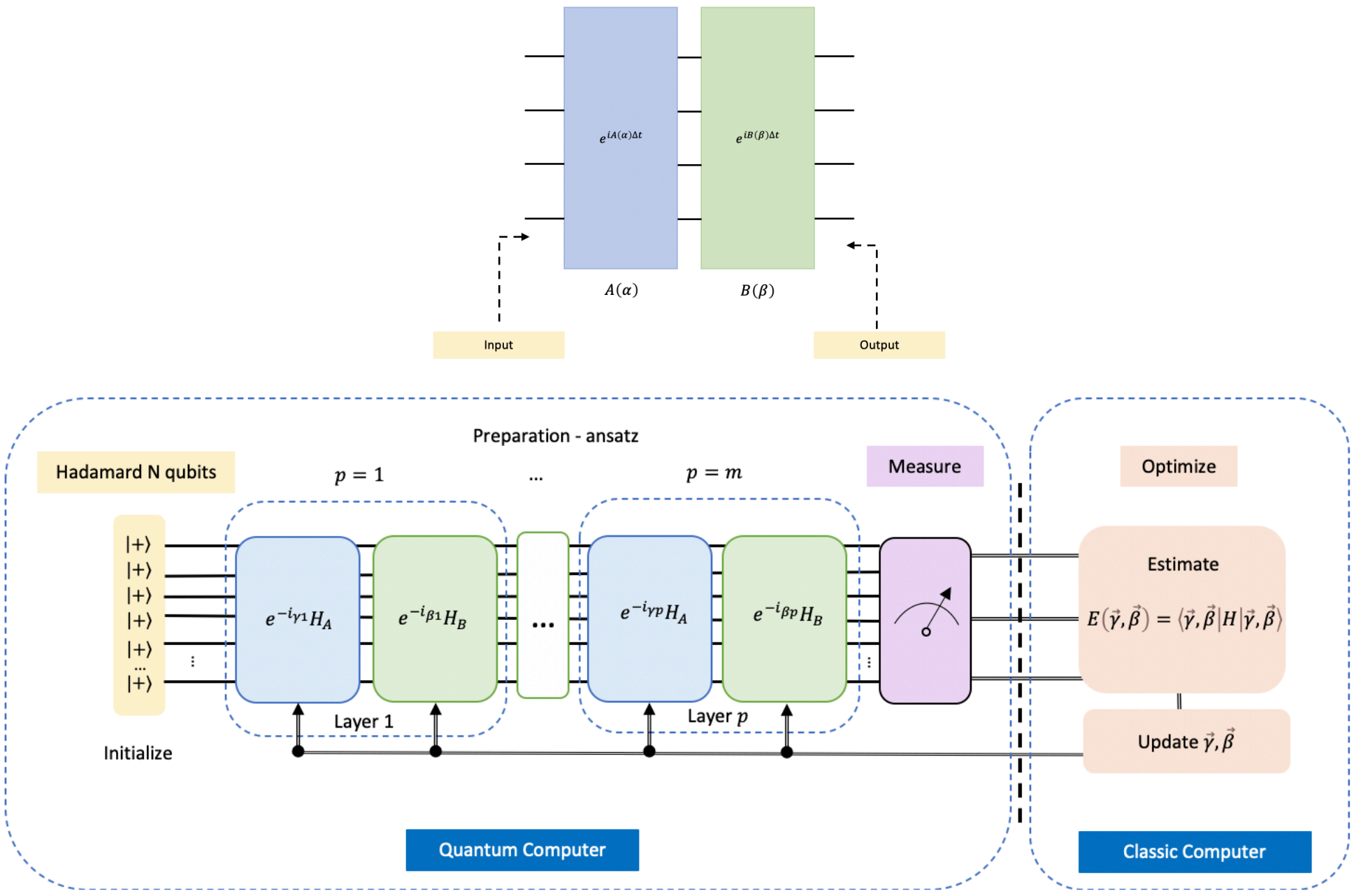

**Fig. 5.** Functional diagram of the Quantum Approximate Optimization Algorithm (QAOA) [19] proposed by Farhi, Goldstone and Gutmann, Blocks $A$ and $B$ are defined through the Hamiltonians $A$ and $B$ that evolve for a short time $\Delta t$. In this case, we will directly map our problem (by its Hamiltonian) in the parameterized circuit. That is, the parameterized circuit is the spitting image of the problem (Hamiltonian objective function). The image below presents a bit in-depth as the QAOA is inside. Where $p$ is the depth of the circuit.

## 2.1 Characteristics of the parameterized quantum circuits

Parameterized quantum circuits or also known as variational circuits are quantum algorithms (from non-electronic design code) that depend on control parameters. Like standard quantum circuits, they consist of three necessary elements. The preparation of a known fixed initial state of a quantum circuit $U(\theta)$; the parametrization by a set of controllable parameters $\theta$ (usually they are angles); and, the measurement of the observable $\widehat{H}$. Then, once the expected value has been calculated, it is iterated with the optimization subroutine. The last two steps are repeated until the optimal $\theta$ parameter $(\hat{\theta})$ is found. In other cases, it may be multiple parameters to minimize. For example, in case of a quantum classifier, it is minimized, both $\theta$ and $x$.

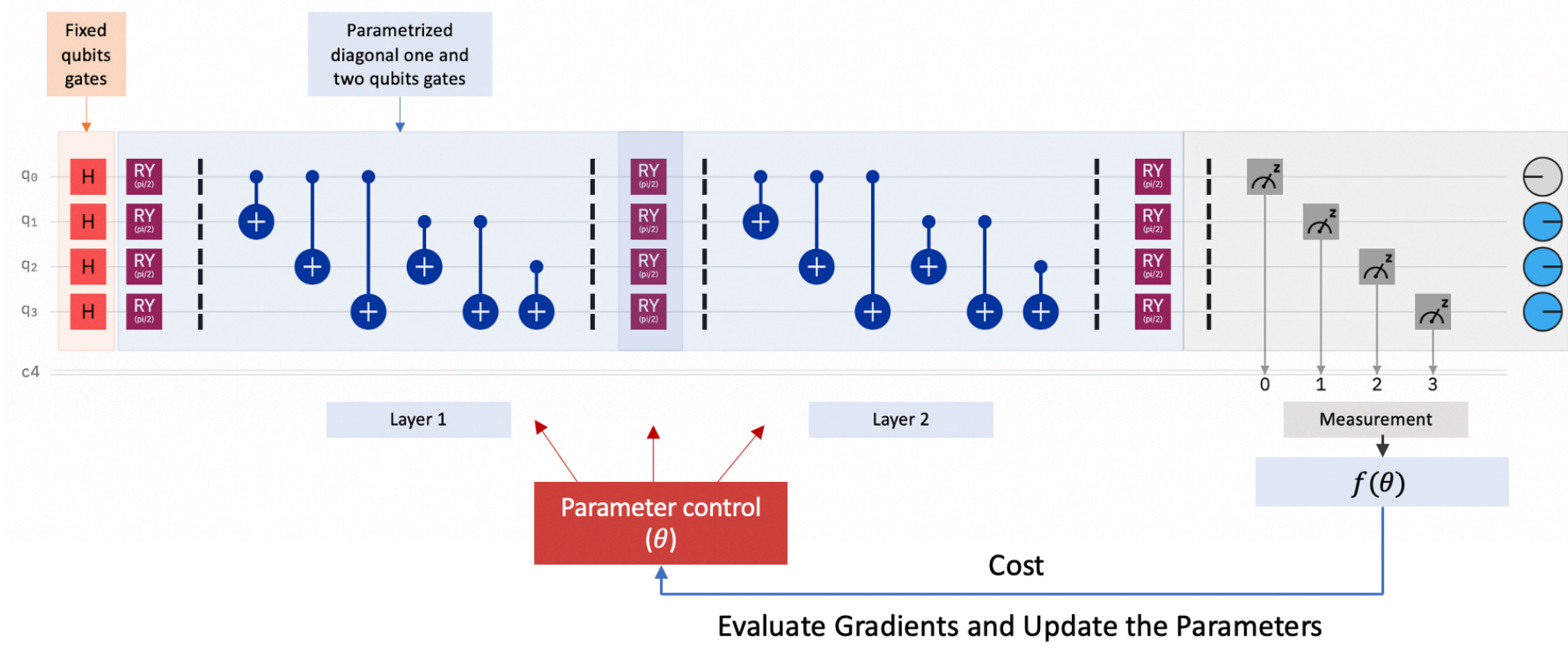

**Fig. 6.** Operation of the variational algorithm updating the parameters through a classical optimization algorithm. This algorithm is known as VQE. What the VQE does when it is run is, leverage on a classic optimizer that varies the parameters, search, either by gradient or not, trying to navigate through that multidimensional parameter space to find the minimum of the operator's expectation given the state from ansatz. What it is achieved with the Hadamard $H$ gate is to make a "base change" to the computational base $Z$.

$$U_3\,(\theta,\phi,\lambda) = \begin{bmatrix} \cos\,({}^{\theta}/_{2}) & -e^{i\lambda}\sin\,({}^{\theta}/_{2}) \\ e^{i\phi}\sin\,({}^{\theta}/_{2}) & e^{i\lambda+i\phi}\cos\,({}^{\theta}/_{2}) \end{bmatrix} \tag{1}$$

$$R_\phi = U_3\,(\theta,\phi,\lambda)|_{\lambda=\phi=0} = U_3\,(0,0,\lambda) = U_1 = \begin{bmatrix} 1 & 0 \\ 0 & e^{i\lambda} \end{bmatrix} \tag{2}$$

It is essential to observe that with the gate $U_3$ (1) any unitary quantum gate [20] can be generated and, above all, with the control of its variables, any PQC can be created that in turn is useful for classification systems. It is also worth mentioning that, in this case, it is observed that the Hamiltonian does not participate in the design of the quantum circuits (Fig. 6.). Only partake in finding the expected value (3).

$$\langle H \rangle = \langle \psi\,(\theta)\,|U^T(\theta)\;\hat{H}U(\theta)\,|\,\psi\,(\theta)\rangle \tag{3}$$

The control parameters $\theta$ of the circuits are adjusted in a feedback loop to optimize this cost function. The computational cost of said feedback in charge of optimizing the control parameters is not so high; therefore, it is a task that classical computers can and usually do efficiently.

From this concept (Fig. 6.), the hybrid computing (Quantum + Classical) of the Noisy Intermediate-Scale Quantum (NISQ) [21] era is born. Which takes advantage of the capacity of quantum computing (PQC and VQE) to solve complex problems, and the experience of classical optimization algorithms (COBYLA, SPSA, etc.) to train variational circuits in which classical algorithms should follow an iterative scheme in search of the best candidates for the control parameters in each loop.

The excellence of the hybrid computing idea in the NISQ era is immense. Since, with a good optimization system, and with a closed-loop system, the noises that characterize this era could be automatically corrected during the optimization process.

Also, with the insertion of information (data) within the variational circuit through the quantum gate $U$, learning techniques can be enhanced. In the next sections, the VQE algorithm and how to add information to the parameterized circuit to create and improve Quantum Machine Learning (QML) [22] will be explored.

# 3 Variational Quantum Eigensolver - VQE

Unfortunately, we're still in the NISQ era because we don't have yet a perfect quantum computer. To compensate for the fact that quantum isn't excellent yet, researchers started developing algorithms that work both quantum and classical parts to solve problems. This area is known as Quantum Machine Learning, and one of the warmest QML algorithms nowadays is the Variational Quantum Eigensolver and the Variational Quantum Classifier. This is because its applications range from finance, biology, scheduling and chemistry. One of the essential characteristics of molecules is its ground state energy. The ground state energy is just the lowest possible energy state that a molecule can be in. The ground state energy of a molecule is really important because it gives us more information about the electron configuration of that molecule.

By varying the experimental parameters in the preparation of the state and calculating the Rayleigh-Ritz ratio [23] using the subroutine in a classical minimization, unknown eigenvectors can be prepared. At the end of the algorithm, the reconstruction of the eigenvector that is stored in the final set of experimental parameters that define the state will be done.

The variational method in quantum mechanics is used, which is a way of finding approximations to the energetic state of lower energy or ground state, and some excited states. This allows to calculate approximate wave functions, such as molecular orbitals and is the basis of this method. It is the variational principle that helps to write the following equation $\langle H \rangle_{\psi(\vec{\theta})} \geq \lambda_i$. With $\lambda_i$ as eigenvector and $\langle H \rangle_{\psi(\vec{\theta})}$ as the expected value. It is clear that the problem that the VQE solves is reduced to finding such an optimal choice of parameters $\vec{\theta}$, that the expected value is minimized and that a lower eigenvalue is found.

$$\langle H \rangle = \langle \psi(\theta) \mid H \mid \psi(\theta) \rangle \tag{4}$$

## 3.1 Noisy VQE for the Optimization problem

Although logic gates are not perfect and have noise, having the variational principle, a golden opportunity can be seen in NISQ devices to have a machine that analyzes the Hilbert vector space. This leads us to forget about the imperfections of the gates and only think of the variational ansatz [9] and that, this ansatz can be analyzed efficiently on a quantum computer. Something a classic computer can't really do. The Fig. 7. summarizes the idea and the basis of Quantum learning [22] [24] [25]; machine learning (ML) on circuit design (PQC).

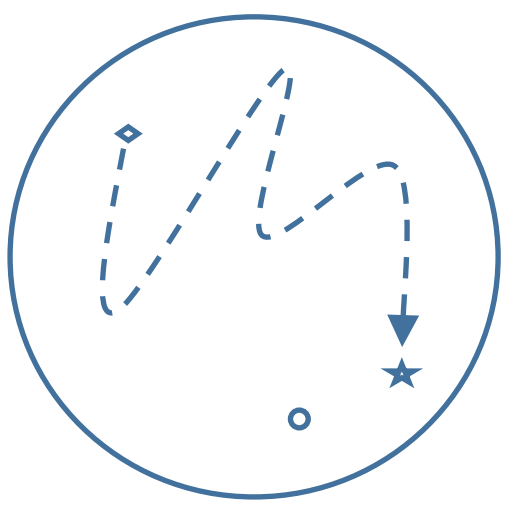

**Fig. 7.** Minimization principle used in Variational Quantum eigensolvers to empower the Quantum Machine Learning.

The term "Noisy Intermediate Scale Quantum" describes the era in which we find ourselves today. Noisy, because the computer still does not offer enough qubits to save for error correction [20]. So, we have imperfect qubits at the physical layer, last but not at least "intermediate scale" because of their small number of qubits.

# 4 PQC as the basis of Machine Learning (ML)

The learning base is closely linked to the optimization process conditioned to a reference, a computing capacity and a space for analyzing the information. In the search for parallelism between the PQC and the learning systems, one only needs to understand how to introduce the data to learn in the variational circuit (Fig. 8.).

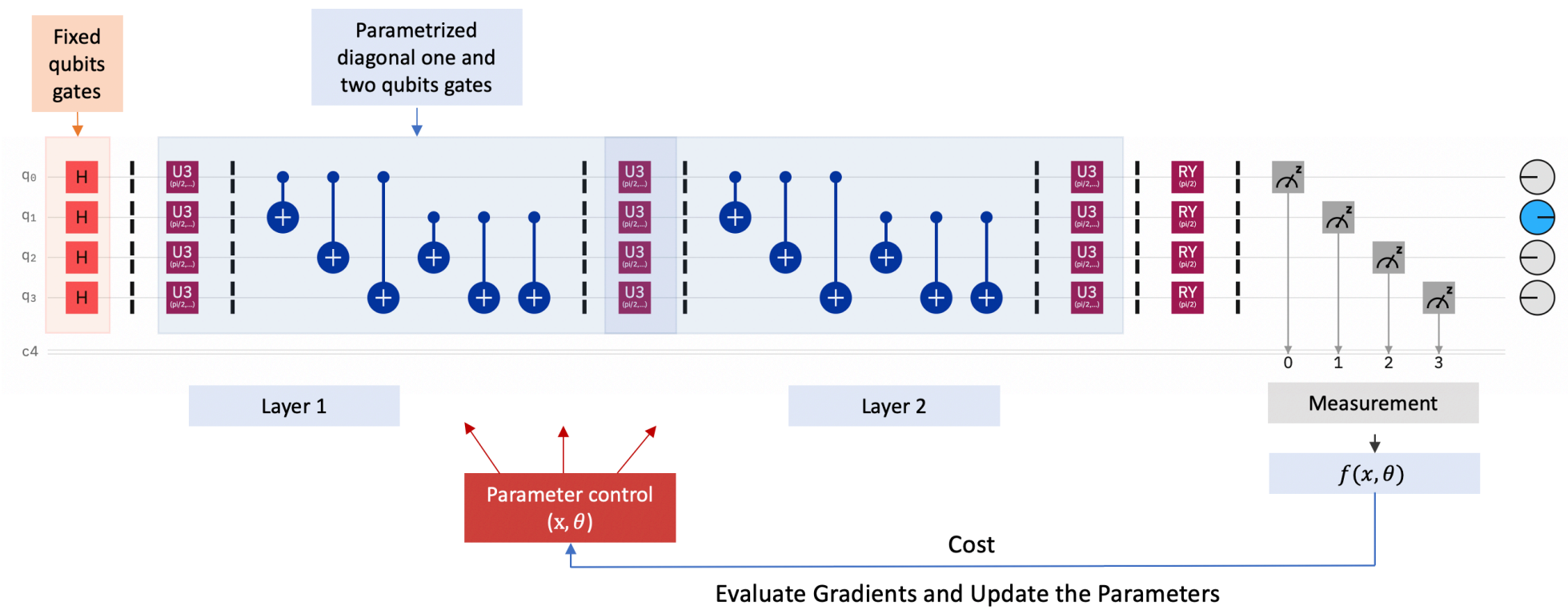

**Fig. 8.** Conceptualization of the PQC as ML for the social workers' problem.

Let the unit gate of equation (1) be $U(x;\theta)$ resulting from the chosen PQC, where it is decided to put the information $(x)$ in the first gate $U_0(x)$ and the variational parameters $\theta$ in the rest of the gates, and one can write the following expression:

$$U(x;\theta) = U_N(\theta_N)U_{N-1}(\theta_{N-1})\cdots U_i(\theta_i)\cdots U_1(\theta_1)U_0(x) \tag{5}$$

$$U(x;\theta)\,|\psi\rangle \tag{6}$$

The variational parameters $\theta$, together with the additional set of non-adaptive parameters $x$ enter the quantum circuit as arguments for the gates of the circuit. This is the process of converting information in the classical domain to quantum information. Equation (6) presents the quantum state associated with the system. Where the non-adaptive gate parameters $(x)$ play the role of data inputs in the quantum machine learning system. Quantum information is converted back to classical information by evaluating the expected value $\widehat{H}$. This powerful concept was first introduced in 2017 by scientists Guillaume Verdon, Michael Broughton and Jacob Biamonte, leveraging on the work done by Fahri in defining the QAOA. Since then, the scientific community is working to make quantum computing a valid option for machine learning [22] [24] [25].
Variational Quantum Computing (VQC) can be viewed as several adaptive electronic circuits in parallel. Each circuit corresponds to an interaction of the optimization algorithm. As if it were several ASICs or FPGAs in parallel, all controlled by the optimization algorithm. In any case, it should be remembered that the choice of the PQC is key.

Let $f(x;\theta)$ be the quantum circuit function and $|\psi_{i-1}\rangle$ be the initial state where $i$ is the position of the gate $U_i$. The initial state can be expressed as $|\psi_{i-1}\rangle = U_{i-1}(\theta_{i-1})\cdots U_1(\theta_1)U_0(x)|0\rangle$. In the same way, the observable $\widehat{H}$ will be given by:

$$\widehat{H}_{i+1} = U_N^{\dagger}(\theta_N)\cdots U_{i+1}^{\dagger}(\theta_{i+1})\widehat{H}U_{i+1}(\theta_{i+1})\cdots U_N(\theta_N) \tag{7}$$

Now it's defined the simplified quantum circuit function $f(x;\theta)$ as:

$$f(x;\theta) = \langle \psi_{i-1} | U_i^{\dagger}(\theta_i) \widehat{H}_{i+1} \; U_i(\theta_i) \; | \psi_{i-1} \rangle \tag{8}$$

Let $\mathcal{M}_{\theta_i}$ be transformation defined as $\mathcal{M}_{\theta_i}(\widehat{H}) = U_i^{\dagger}(\theta_i)\widehat{H}U_i(\theta_i)$, by simplifying equation (8), one arrives at the following equation:

$$f(x;\theta) = \langle \psi_{i-1} | \mathcal{M}_{\theta_i}(\widehat{H}_{i+1}) \; | \psi_{i-1} \rangle \tag{9}$$

**Table 1.** One schedule of patient visits without any association with social workers. Where $U_1$ to $U_5$ are the patients (users) and equal to the variable $i$ or $j$ of the mathematical formulation.

| | M | T | W | T.H. | F |
|---|---|---|---|---|---|
| 9:00 – 10:00 | $U_1$ | $U_1$ | $U_1$ | $U_1$ | $U_1$ |
| 9:30 – 10:30 | $U_4$ | | | | |
| 10:15 – 11:15 | | | $U_4$ | | |
| 11:30 – 12:30 | | | | | $U_5U_4$ |
| 11:45 – 12:45 | $U_5$ | | | | |
| 12:00 – 13:00 | $U_2$ | | | | |
| 14:45 – 15:45 | | $U_2$ | | | |
| 15:00 – 16:00 | | | | $U_3$ | |
| 15:15 – 16:15 | $U_3$ | | | | |
| 15:45 – 16:45 | | | $U_3$ | | |
| 16:00 – 17:00 | | | $U_2$ | | |
| 16:30 – 17:30 | | $U_3$ | | | |
| 17:00 – 18:00 | | | | | $U_3$ |

### 4.1 Calculation of the gradient of the chosen PQC

Consider a quantum computer with parameterized gates (derived from equation (1)) of the form:

$$U_i(\theta_i) = e^{\left(-i\frac{\theta_i}{2}\widehat{\sigma_i}\right)} \tag{10}$$

With $\sigma_i = \sigma_i^{\dagger}$, a Pauli operator. The following equation gives the gradient of this unit:

$$\nabla_{\theta_i} U_i(\theta_i) = -\frac{i}{2}\widehat{\sigma_i}U_i(\theta_i) = -\frac{i}{2}U_i(\theta_i)\widehat{\sigma_i} \tag{11}$$

Now substituting equation (11) in the quantum circuit function $f(x;\theta)$:

$$\nabla_{\theta_i} f(x;\theta) = -\frac{i}{2}\langle\psi_{i-1}|U_i^{\dagger}(\theta_i)\left(\sigma_i\hat{H}_{i+1} - \hat{H}_{i+1}\sigma_i\right)U_i(\theta_i)|\psi_{i-1}\rangle \\ =_{|[X,Y]} -\frac{i}{2}\langle\psi_{i-1}|U_i^{\dagger}(\theta_i)\ [\sigma_i,\hat{H}_{i+1}]U_i(\theta_i)\ |\ \psi_{i-1}\rangle \tag{12}$$

The following mathematical identity is now used for daily trips involving Pauli operators. It is recalled that the computational basis (in Qiskit) is that of the Pauli operator in dimension $Z$.

Using the commutation with the Pauli operators ($\sigma_i$), within the quantum circuit, one arrives at the equation (13),

$$[\hat{\sigma}_i,\hat{H}] = -i\left(U_i^{\dagger}\left(\frac{\pi}{2}\right)\hat{H}U_i\left(\frac{\pi}{2}\right) - U_i^{\dagger}\left(-\frac{\pi}{2}\right)\hat{H}U_i\left(-\frac{\pi}{2}\right)\right) \tag{13}$$

$$\nabla_{\theta_i} f(x;\theta) = \frac{1}{2}\left\langle\psi_{i-1}\left|U_i^{\dagger}\left(\theta_i+\frac{\pi}{2}\right)\hat{H}_{i+1}\ U_i\left(\theta_i+\frac{\pi}{2}\right)\right|\psi_{i-1}\right\rangle \\ -\frac{1}{2}\langle\psi_{i-1}|U_i^{\dagger}\left(\theta_i-\frac{\pi}{2}\right)\hat{H}_{i+1}\ U_i\left(\theta_i-\frac{\pi}{2}\right)|\ \psi_{i-1}\rangle \tag{14}$$

$$\nabla_{\theta_i} f(x;\theta) = \frac{1}{2}\left[f\left(x;\theta_i+\frac{\pi}{2}\right) - f\left(x;\theta_i-\frac{\pi}{2}\right)\right] \tag{15}$$

At this point, knowing how to define the quantum classifier, how to optimize the variational parameters and knowing the output of the variational circuit, the chosen PQC is confirmed. Now it only remains to see in detail the proposed problem and the approach to solve it.

## 5 Proposed problem

Let $n$ be the number of patients (users) and considering a weekly calendar of visits for each of them. The objective is to find an optimal meeting's calendar, which minimizes the cost of time travel; hence, money and maximizes the number of visits to the patients in a work schedule. And if there are changes from the social workers or they hire or fire, new workers or any change in patients, the system must adapt dynamically (see Fig. 9.). In the end, it assigns social workers to the group with the resultant optimal hours. A change of organization and association of patients and social workers according to an optimal calendar is only allowed after three months of the last generation of the optimal calendar. A patient is not allowed to change social workers before three months or when the degree of optimization of the social workers' calendar is low. This means that a social worker is assigned to a patient for three months, and before this date, it is not allowed to change it. This avoids making the patient dizzy with changes in social workers.

In this case study, the daily schedule (see Table. 1.), is set at 7 hours, and the distance between patients is at least 15 minutes.
The experiment's scenario and its design are representatives because the only difficulty that can be added here, for this combinatorial optimization problem, is the number of patients, social workers and some restrictions. These difficulties have to do directly with the computational cost (not with the formulation/algorithm) where quantum computing is called to be more efficient [21]. Hoping to have access to a more powerful computer, the test is limited to a 15-qubits computer (*ibmq_16_melbourne*); the most potent public quantum computer. However, we can simulate the algorithm in simulators with a higher number of qubits.

In this article, $n = 5$ ($N = n(n-1) = 20$ qubits) because the experiments from the problem of this class are among the most difficult in the test bench. This fact will be reviewed in the discussion chapter.

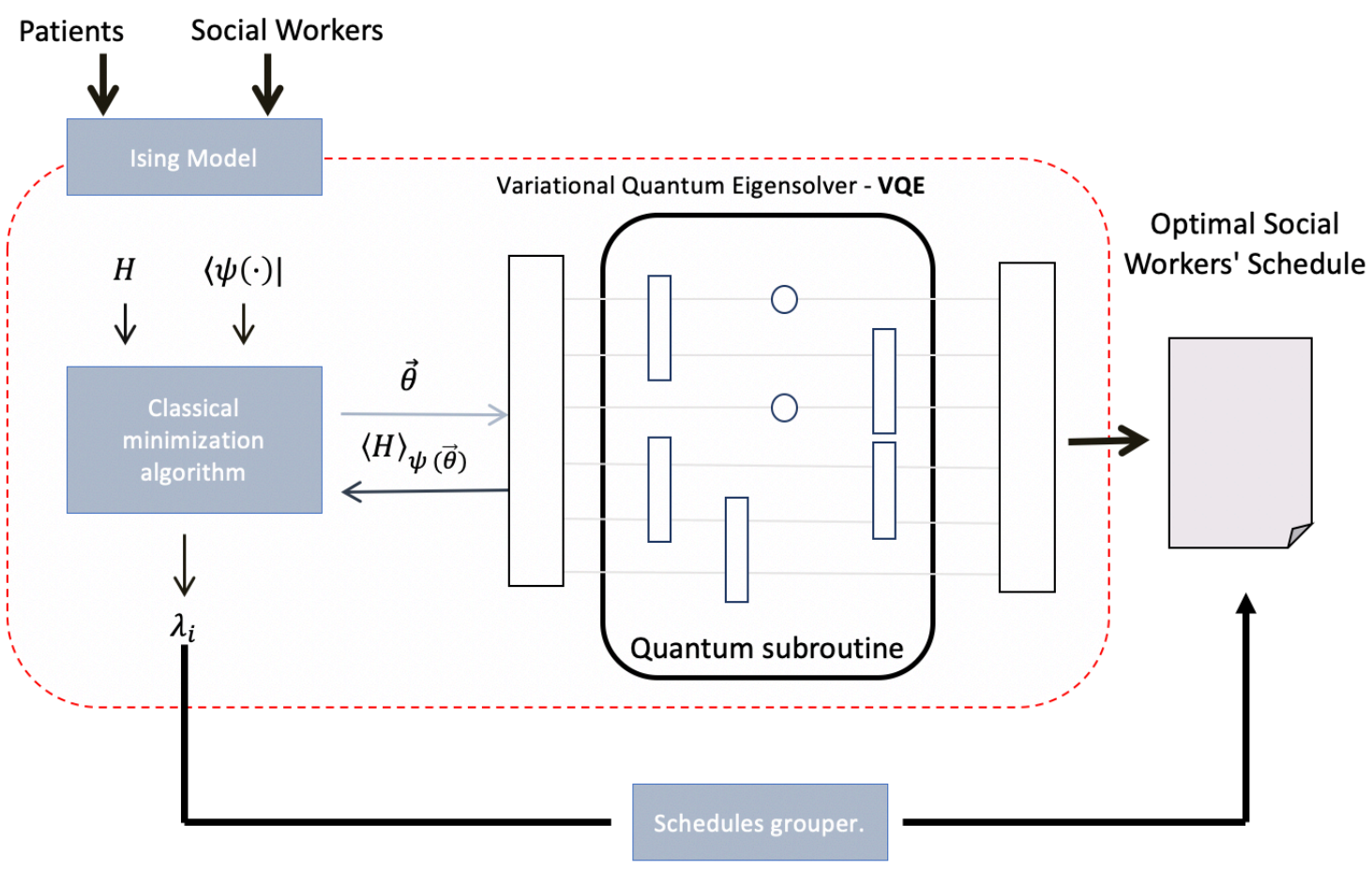

**Fig. 9.** Using the Variational-Quantum-Eigensolver (VQE) as Quantum Machine Learning to creates an Intelligent social workers schedule problem solver.

## 6 The approach to solve the problem

To date, what has been done is to validate the formulation of the problem of social workers in previous works. What is being done is based on creating a parameterized quantum circuit with a relative depth that allows generating the best possible state vector in the Hilbert vector space. And use the VQE and its derivatives, to calculate the energy of the tracking state associated with the Hamiltonian of social workers and iterate through a loop with the classical optimizer until you find the minimum eigenvalue (the ground state), to calculate the best optimal time for social workers.

Now, let's imagine that one of the users or patients undergoes some changes. By having a restriction that it is not possible to change the social worker related to a given patient during the first three months from the generation of the visit calendar, two strategies are defined to calculate a best optimum without the need to compute from scratch.

The first strategy is based on having an ordered queue from optimal to the least optimal user. That is, to have an order of users who can still make more visits or not depending on their contract and patient assignment. If new patients sign up, they will be assigned to social workers who can do a few more hours, based on a priority rule. In case of cancellation, the system will wait without recalculating anything until three months have elapsed since the last calculation of the best optimal or, depending on the degree of degradation of the visiting hours.

The second strategy that is implemented in this article to optimize a best-recovered case is to configure the VQE / VQC to calculate from the last optimal point, as reflected in figure (10). If the previous optimal solution is at the point of the star, it would not be necessary to recalculate everything, but to go from the star to the next optimal time that is represented in the figure by the red point. Always after retrieving the information (data that allows to reformulate the input data of the already calculated solution) of the last optimal point.

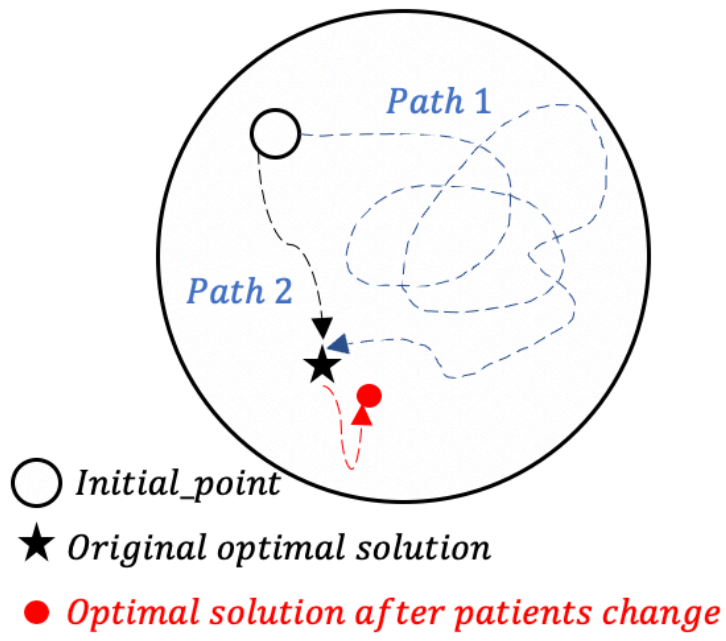

**Fig. 10.** Minimization principle used in Variational Quantum eigensolvers to empower the Quantum Machine Learning. At the moment we execute the VQE, using the optimizer that varies the parameters, it searches by navigating through the multidimensional parameter space to find the minimum of the operator's expectation given the state of our ansatz.

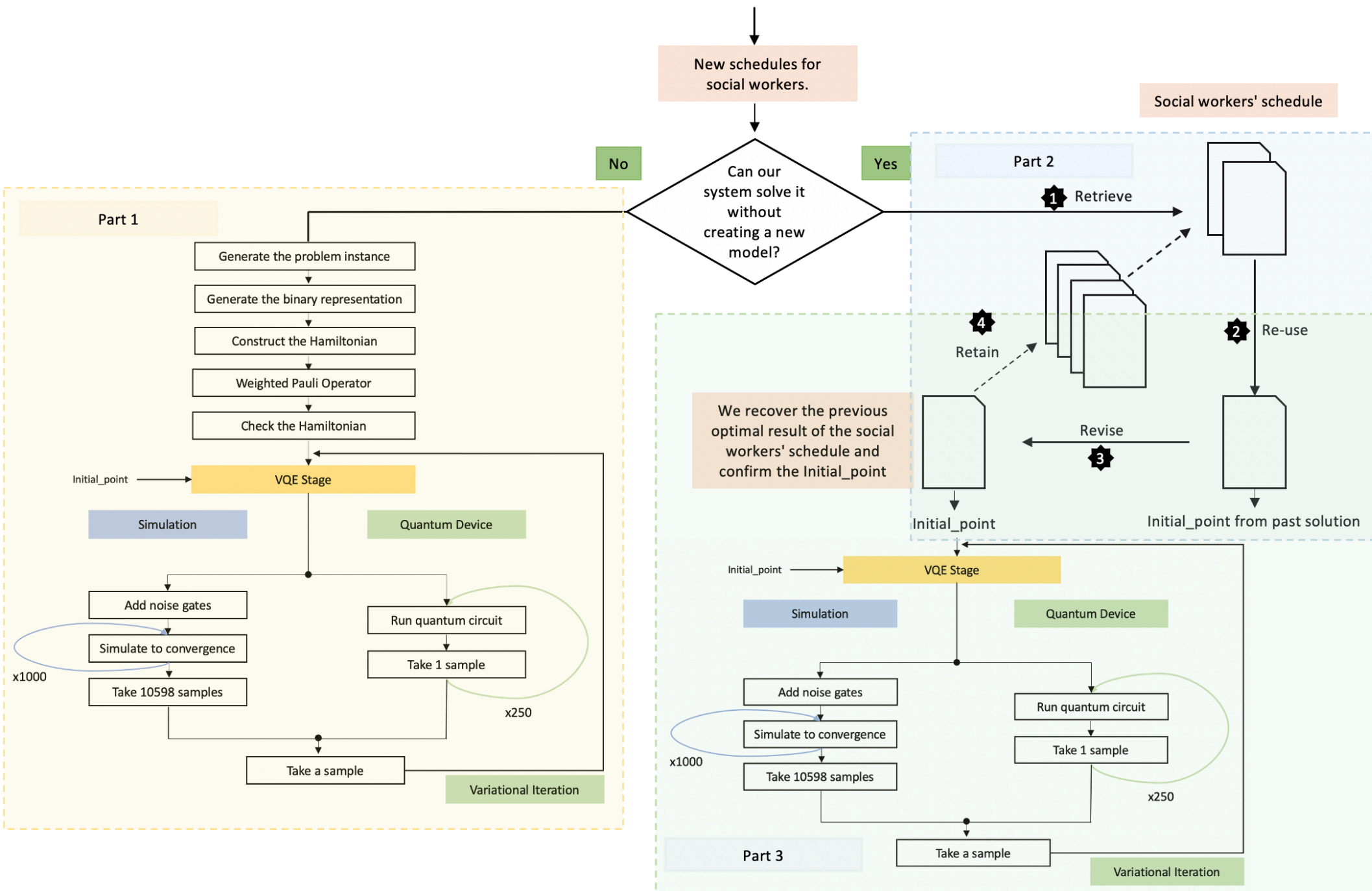

**Fig. 11.** Block diagram of the steps our algorithm follows to solve the social workers' problem. Part 1: The Top-Down quantum resolution of the quantum problem. This is the specific resolution range of any quantum routing and scheduling algorithm. This branch will run a lot at startup (at system startup) in the case of not having enough memory from the previous cases. Part 2: Applying Case-Based Reasoning retrieving an existing solution, but without optimizing the solution retrieved. In this case, it is retrieved last optimal schedule of the social workers (exploring in the exit space and then in the space of the problem statement; at the entrance). And it is observed if the case is the best by approximation that can be guaranteed without the need to optimize the recovered solution. Part 3: The resolution based on the recovery of an existing solution (recovery of the case by an exploration of the statement and solution space), but with the need to optimize the recovered solution.

Therefore, the algorithm (Fig. 11.) is broken down into two key parts, the last (third) being a generalization of the second point:

1. The Top-Down from the scratch resolution: this is the specific resolution range of any quantum routing and scheduling algorithm (see previous works). In the case of not having enough memory of the earlier schedules, this branch will be executed much more to start the system.

2. The resolution based on the recovery of an existing solution, but without the need to optimize the recovered solution: in this case, it is recovered the optimal schedule of the social workers (exploring in the output space and then in the space of the problem statement; at the entrance), which approaches the solution of the new problem and guarantees a new solution without the need to optimize said recovered solution.
3. The resolution based on the recovery of an existing solution (recovery of the case by an exploration of the statement and solution space), but with the need to optimize the recovered solution.

It is worth mentioning that, during the transitory regime (there is still not enough memory of cases), part 1 will be executed more. Contrary to the transitory regime, the permanent regime, which is that of regular operation (part 2 and 3). Where there are several cases that have already been solved (memory of cases).

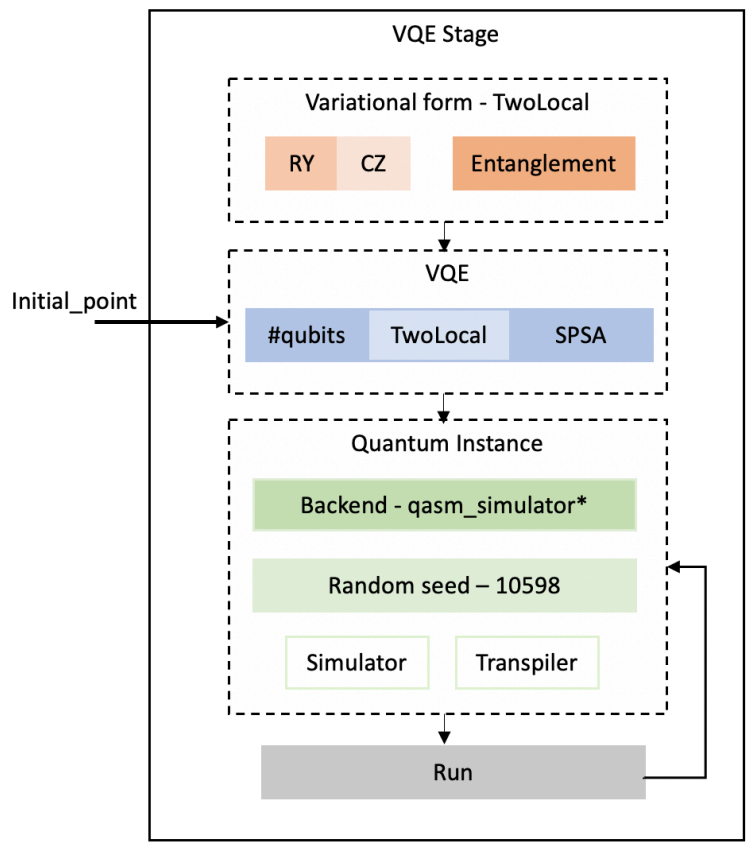

**Fig. 12.** Block diagram of the VQE that we use in parts 1 and 3 of the algorithm to solve the social workers' problem.

### 6.1 Steps to follow to solve the problem

What the designed algorithm does is described below. Given some new schedules for social workers, the system looks in the space of the entry hours, to see if there are similarities with the schedules already resolved previously. It also verifies that the changes in the new schedules do not imply recalculating the specific solution through the Ising model. If it is not necessary to recalculate everything, the algorithm will follow the following steps: first, it will contextualize the new problem. Second, it will search for a previously solved similar problem, retrieve the solution, adapt the solution to the current issue, and finally, it will verify the solution and store the newly solved problem to solve future problems.
In the case that it is a new schedule in its entirety, what the algorithm will do is the model the problem and look for its translation in quantum by finding its Ising model. Once having the Ising model (the qubits and offset), the problem will be solved by executing the VQE from the last *initial_point* (Fig. 12). The *initial_point* is a configuration parameter of the VQE from which the optimizer starts. This point is useful for cases in which there is a certain similarity with the new problem; no need to recalculate everything. That is, it is not necessary to search the entire solution space again, but to start from the last optimal point. If the result is queued of the problem that gave the optimal solution (the initial point (*initial_point*), the configuration parameters and the variables) the initial problem can be recovered. Figures (13 and 14) (see figure (12)) represent the block diagram that describes the resolution of the social workers' problem using Case-Based Reasoning.

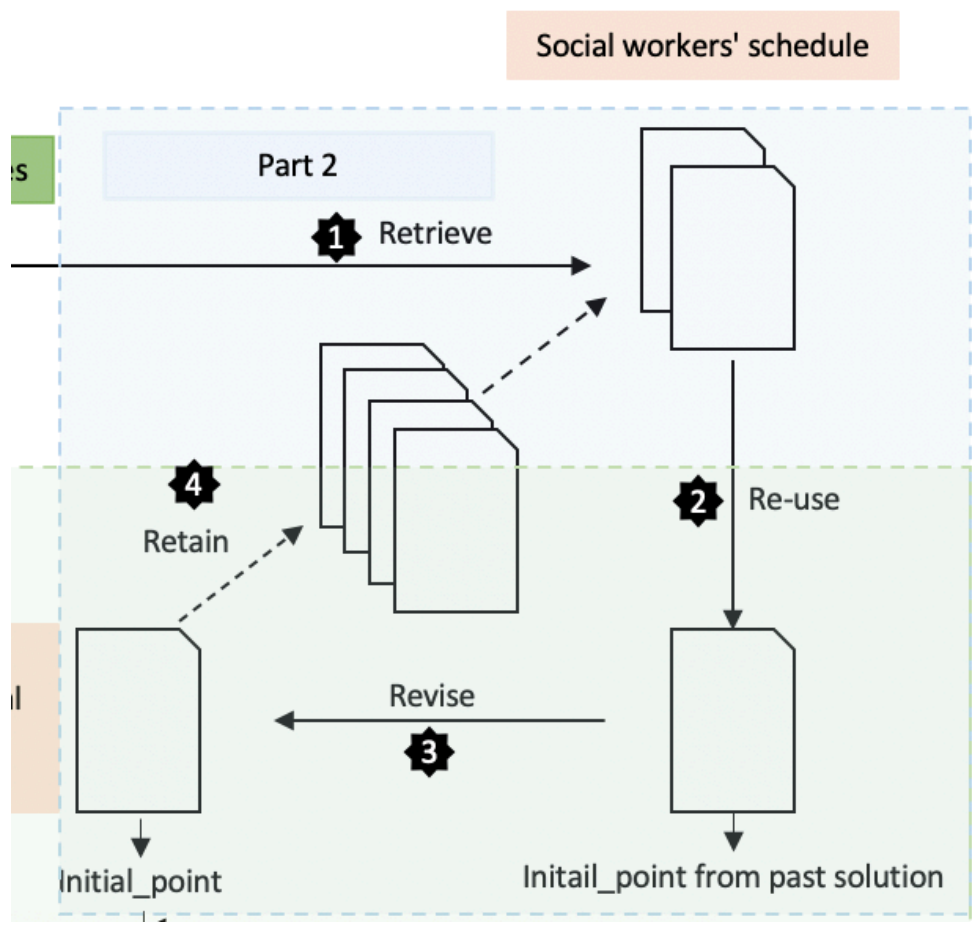

**Fig. 13.** Case-based system block diagram

Next, we will explain the four key steps of the algorithm based on case reasoning:

1. The first thing the algorithm will do is recover the cases already solved to evaluate the new changes in the schedules of the social workers' problem. From the best matching case, the base of the cases is searched, and an approximate solution is retrieved.
2. The algorithm will adapt the case and reuse the retrieved solution to adapt it to the new problem better.
3. The algorithm will review and evaluate the retrieved solution that can be done before the solution is applied to the problem or after the solution has been used to it. In any case, if the achieved result is not satisfactory, the recovered solution must be adapted again or return to point 1.
4. Finally, the algorithm will update the memory of the cases with the correct solution.

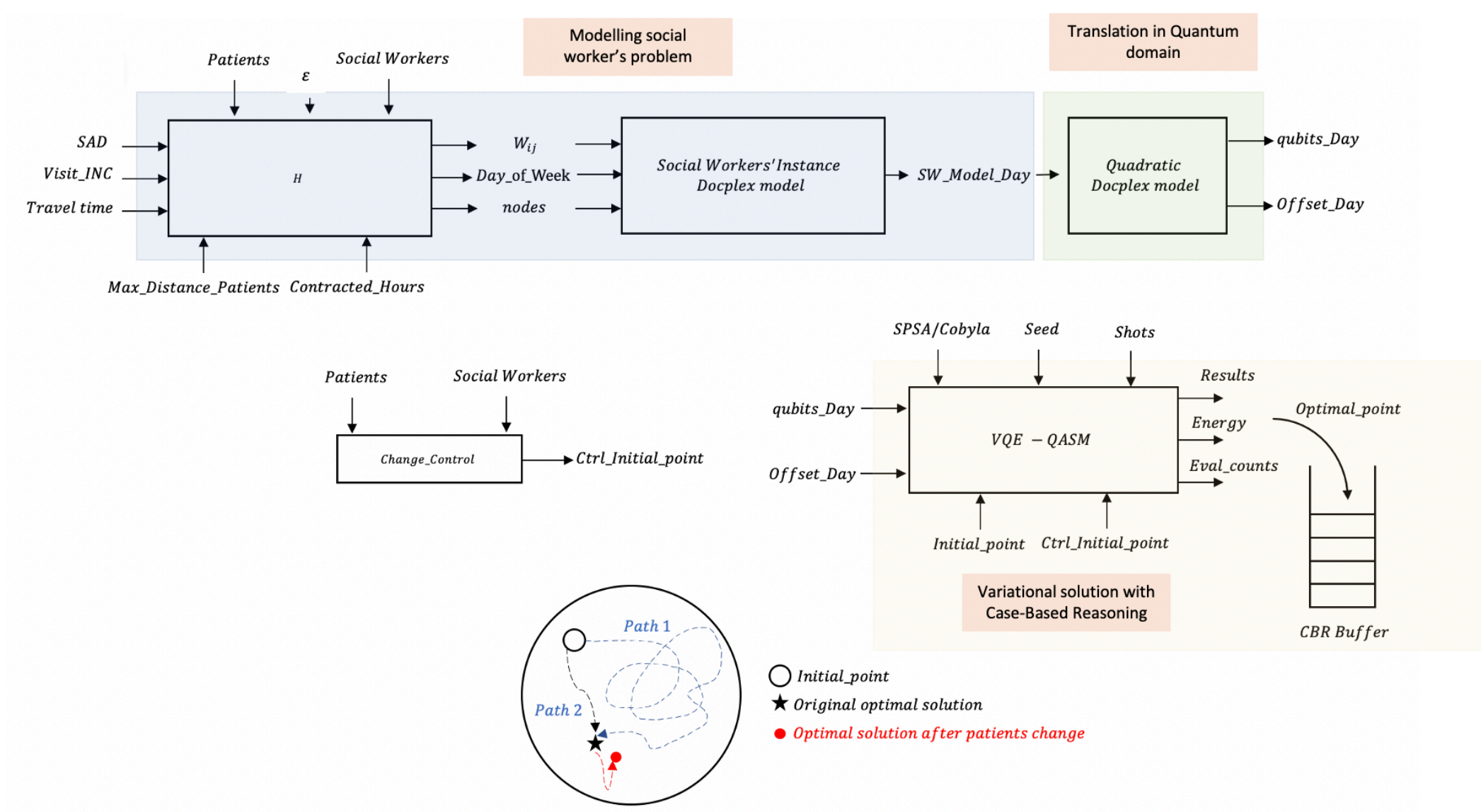

**Fig. 14.** Block diagram of the social workers' problem algorithm with a case-based reasoning system.

# 7 Social worker formulation

Let's recall the formulation of the problem of social workers.
Let $G = (V, E, c)$ be a complete graph directed with $V = \{0,1,2,..,n\}$, as the set of nodes and $E = \{(i,j) : i,j \in V, i \neq j\}$ as the set of arcs, where node 0 represents the central. For a team of $k$ social workers with the same capacity $q$ and $n$ remaining nodes that represent geographically dispersed patients.
Let $W_{ij}$ be the growing travel cost that is associated with each arc $(i,j) \in E$. Let's consider that the $W_{ij}$ distances are symmetrical to simplify the calculation and let $x_{ij}$ be the decision variables of the paths between two consecutive patients.

The goal will be to minimize the following objective function:

$$\sum_{i=1}^{n} \sum_{j=1, i \neq j}^{n} W_{ij} x_{ij,} \tag{16}$$

Subject to:

$$\sum_{i=1}^{n} \sum_{j=1 i \neq j}^{n} x_{ij} = 1, \quad \forall i \in \{1, \dots, n\} \tag{17}$$

$$\sum_{i=1}^{n} \sum_{j=1 i \neq j}^{n} x_{ji} = 1, \qquad \forall j \in \{1, \dots, n\} \tag{18}$$

$$\sum_{i=1}^{n} di \sum_{j=1 i \neq j}^{n} x_{ij} \leq q, \forall i,j \in \{0, \dots, n\}, i \neq j \tag{19}$$

$$\sum_{i=1}^{n} x_{0i} = K, \quad \forall j \in \{1, \dots, l\} \tag{20}$$

$$\sum_{j=1}^{n} x_{j0} = K, \forall i \in \{1, \dots, l\} \tag{21}$$

$$\sum_{i=1}^{n} x_{ih} - \sum_{j}^{n} x_{hj} = 0 \quad \forall h \in \{1, \dots, n\} \tag{22}$$

$$x_{ij} \in \{0,1\}, \forall i,j \in \{0, \dots, n\}, i \neq j \tag{23}$$

The objective function (16) minimizes the total travel savings given the new cost function with the time window we formulate. The restrictions of equation (17) impose that exactly the arcs $k$ leave the plant; (20) and (21) are the restrictions on the degree of entry and exit of the social worker from the central workplace. With the restrictions (22) we ensure that the solution does not contain a sub-route and the last restrictions (23) are obligatory and define the type of linear programming.
Up to this point, the mathematical formulation of equations (16) to (23) represents a conventional VRP. To solve the social workers' problem as a scheduling problem, we may need a time variable. This is so important to finish mapping the cited problem.

## 8 Formulation proposed

Let's recall the proposed formulation expressed by equations 24 and 25 to solve this problem as routing and scheduling task.

$$W_{ij} = d_{ij} + g(t_{ij}) \tag{24}$$

$$g(t_{ij}) = \varepsilon \frac{(\tau_{i-}\tau_j)^2}{d_{max} - d_{min}} \tag{25}$$

Where $W_{ij}$ is the weight time window function, $d_{ij}$ is the distance between the patient $i$ and the next $j$ and $g_{ij}$ is the time window's function. With $g_{ij}$ as a non-negative function that it is mapped on a quadratic function to weigh extremal distances (shortest in relation to the greatest ones). Let's consider that the initial weight function $W_{ij} = d_{ij}$ is a distance function because one wants to make $g_{ij}$ behave like $d_{ij}$, and thus, be able to take full advantage of the behaviour of the initial objective function.
Let $\varepsilon$ be positive and represents a weighted degree parameter of the time window function; $\tau_i$ is the starting worker time of a slot of time for patient $i$ and $\tau_j$ for the patient $j$. With $d_{max}$ as the maximum distance between all patients and $d_{min}$ the minimum one. Hence, let's define the non-negative time windows $T_{ij} = (\tau_{i-}\tau_j) > 0$.

The simplified objective function subjected to the restrictions in Hamiltonian form for the schedule optimization problem is as follows:

$$\begin{aligned} H = \sum_{ij\in E}(d_{ij} + \varepsilon \frac{(\tau_{i-}\tau_j)^2}{d_{max} - d_{min}})x_{i,j} + A\sum_{i=1}^{n}(1 - \sum_{j\in\delta(i)^+}^{N} x_{i,j})^2 \\ + A\sum_{i=1}^{n}(1 - \sum_{j\in\delta(i)^-}^{N} x_{ji})^2 + A(k - \sum_{i\in\delta(0)^+}^{N} x_{0,i})^2 + A(k - \sum_{j\in\delta(0)^+}^{N} x_{j,0})^2 \end{aligned} \tag{26}$$

Where $A$ is a free parameter such that $A > max\left(d_{ij} + \varepsilon \frac{(\tau_{i-}\tau_j)^2}{d_{max}-d_{min}}\right)$.

Now the docplex is applied, or the related Ising model is found for the Hamiltonian equation (26). As stated above, this task requires specialized knowledge of matrix vectorization in a vector using the Kronecker product to express multiplication as a linear transformation in matrices. In this article, we use the docplex to get the compact Ising Hamiltonian ready to compute in NISQ.

The result can have the following mathematical form:

$$\begin{aligned} H = A\sum_{i=1}^{n}[(e_i\otimes\mathbb{I}_n)^2Z^2 + [v_i^T]^2Z^2] + w - 2A\sum_{i=1}^{n}[(e_i\otimes\mathbb{I}_n^T) + v_i^T] - 2Ak\,[(e_0\otimes 1_n)^T \\ + v_0^T] + 2An + 2Ak^2 \end{aligned} \tag{27}$$

With

$$W_{ij} = d_{ij} + \varepsilon \frac{(\tau_{i-}\tau_j)^2}{d_{max} - d_{min}} \tag{28}$$

## 9 Results - experimentation

In this section, a large part of the results of experimentation combining quantum computing with case-based problem-solving techniques are presented. Before testing the algorithm on the quantum computer, IBM Quantum Experience (IBMQ)[1] is used for both the design part and the code lab in order to test and tune our code.

The algorithm mapped in VQE and VQC is tested in *ibmq_16_melbourne v1.0.0 and ibmq_qasm_simulator v0.1.547* with COBYLA[2] and SPSA[3], as the classical optimizers. Fig. 24 to 30 show the results of the algorithm once it is run under the IBMQ.

As a consequence, the problem is solved by creating a quantum circuit for each shot, and the best circuit will be the one that optimizes the problem for social workers' problem.

**Table 2.** Table of the five scenarios taking into account the number of social workers patients

| # Patients | # Social Workers |
|---|---|
| 3 | 2 |
| 4 | 3 |
| 5 | 2 |
| 5 | 3 |
| 5 | 4 |

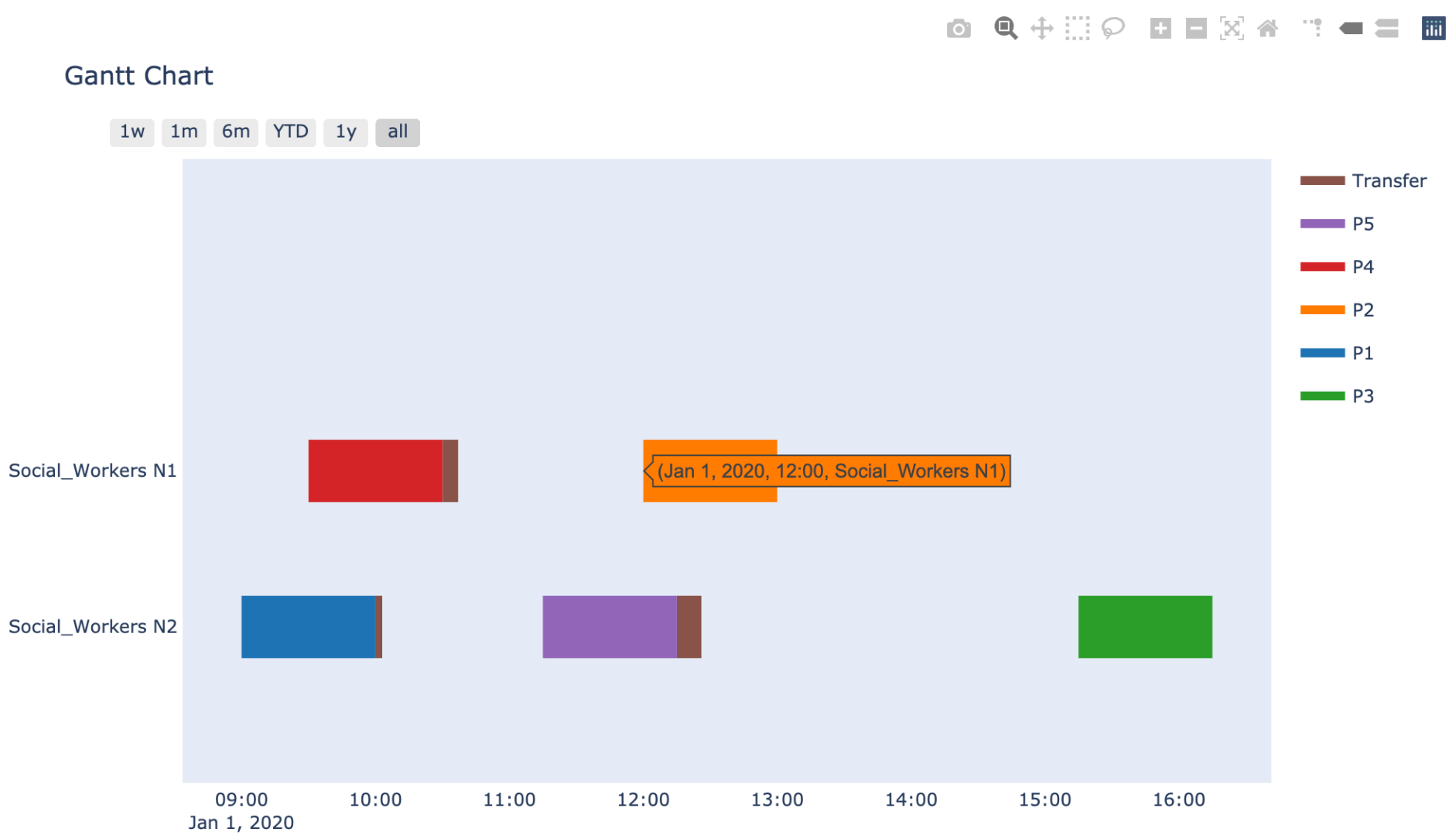

**Fig. 15.** Result of the social workers' problem algorithm. In this case, it is used the 32-qubit *ibmq_qasm_simulator* and visualize the Gantt of the result of the five patients and two social workers with ploty[4]. The figure shows the window of transfer of a visit and another that the coordinator of social workers would see. In this experiment, due to the small number of visits and the considerable distance between visits, the role of the transfer window is not much appreciated.

---

[1] https://quantum-computing.ibm.com/
[2] https://docs.scipy.org/doc/scipy/reference/optimize.minimize-cobyla.html
[3] https://www.jhuapl.edu/SPSA/
[4] https://plotly.com/

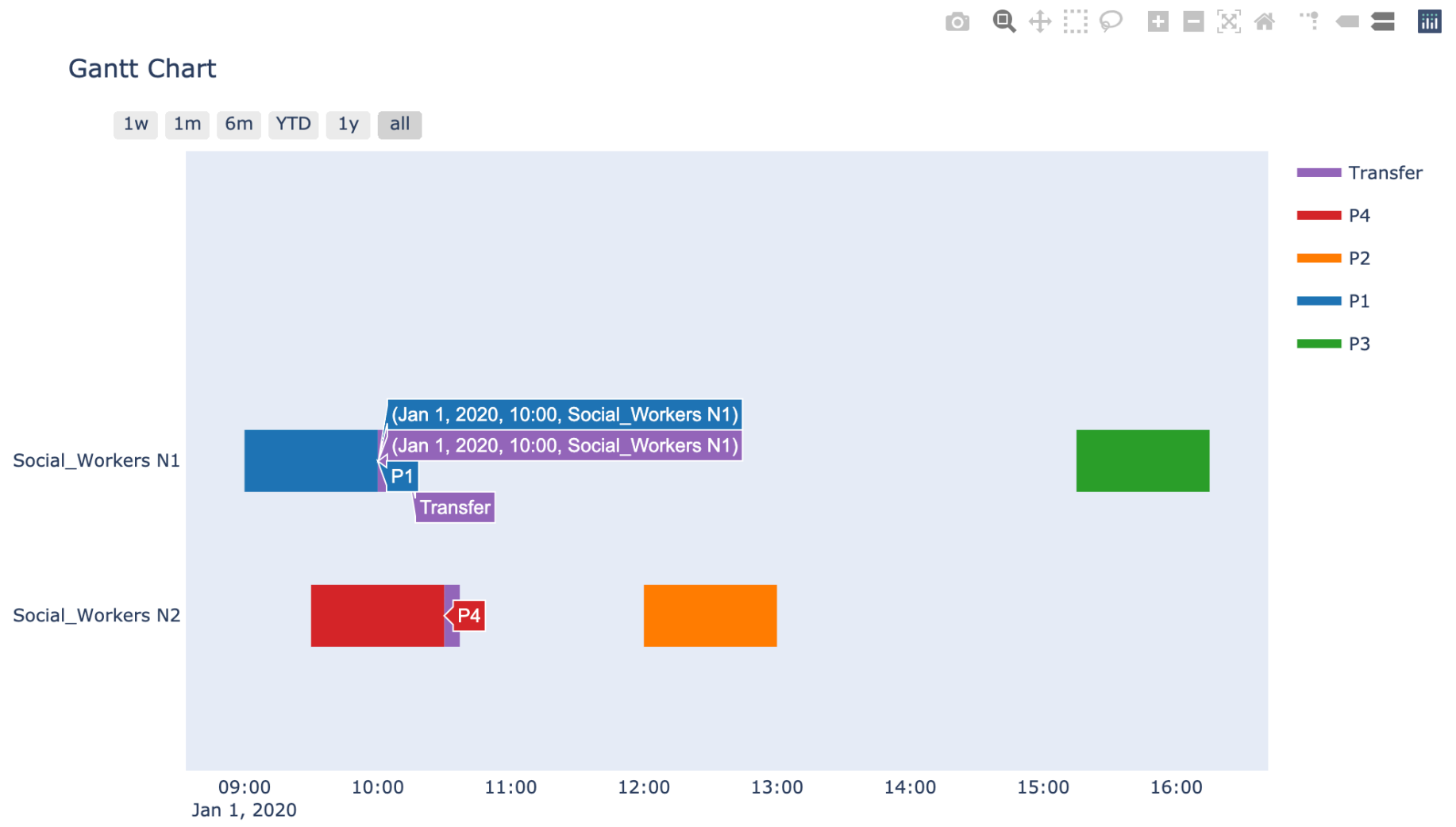

**Fig. 16.** Result of the social workers' problem algorithm. In this case, it is used the *ibmq_16_melbourne* and visualize the Gantt of the result of the four patients and two social workers with ploty. The figure shows the window of transfer of a visit and another that the coordinator of social workers would see. In this experiment, due to the small number of visits and the considerable distance between visits, the role of the transfer window is not much appreciated.

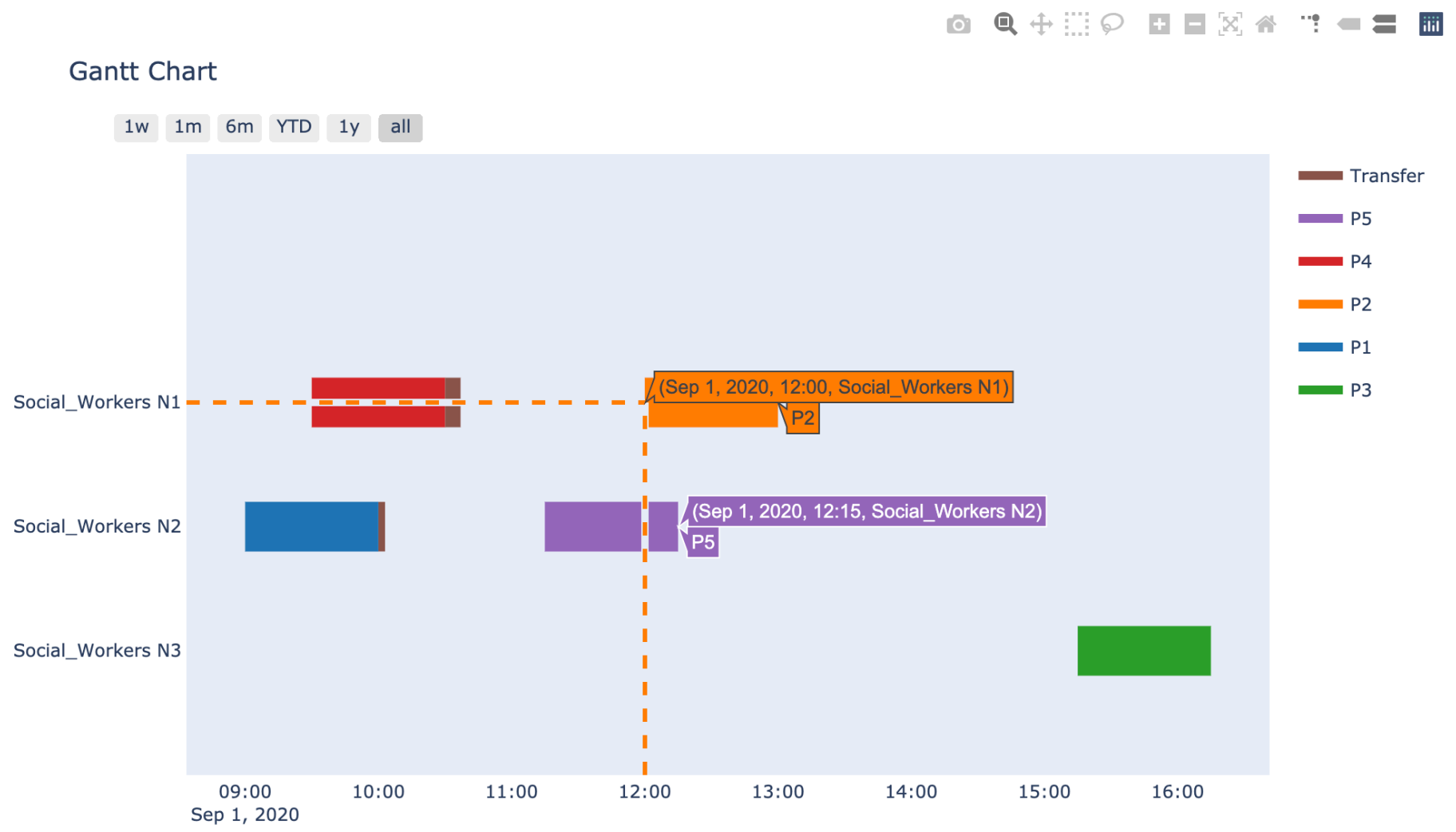

**Fig. 17.** Result of the social workers' problem algorithm using the *ibmq_qasm_simulator*. In this scenario, it is proposed the hypothesis of 5 patients and three social assistants.

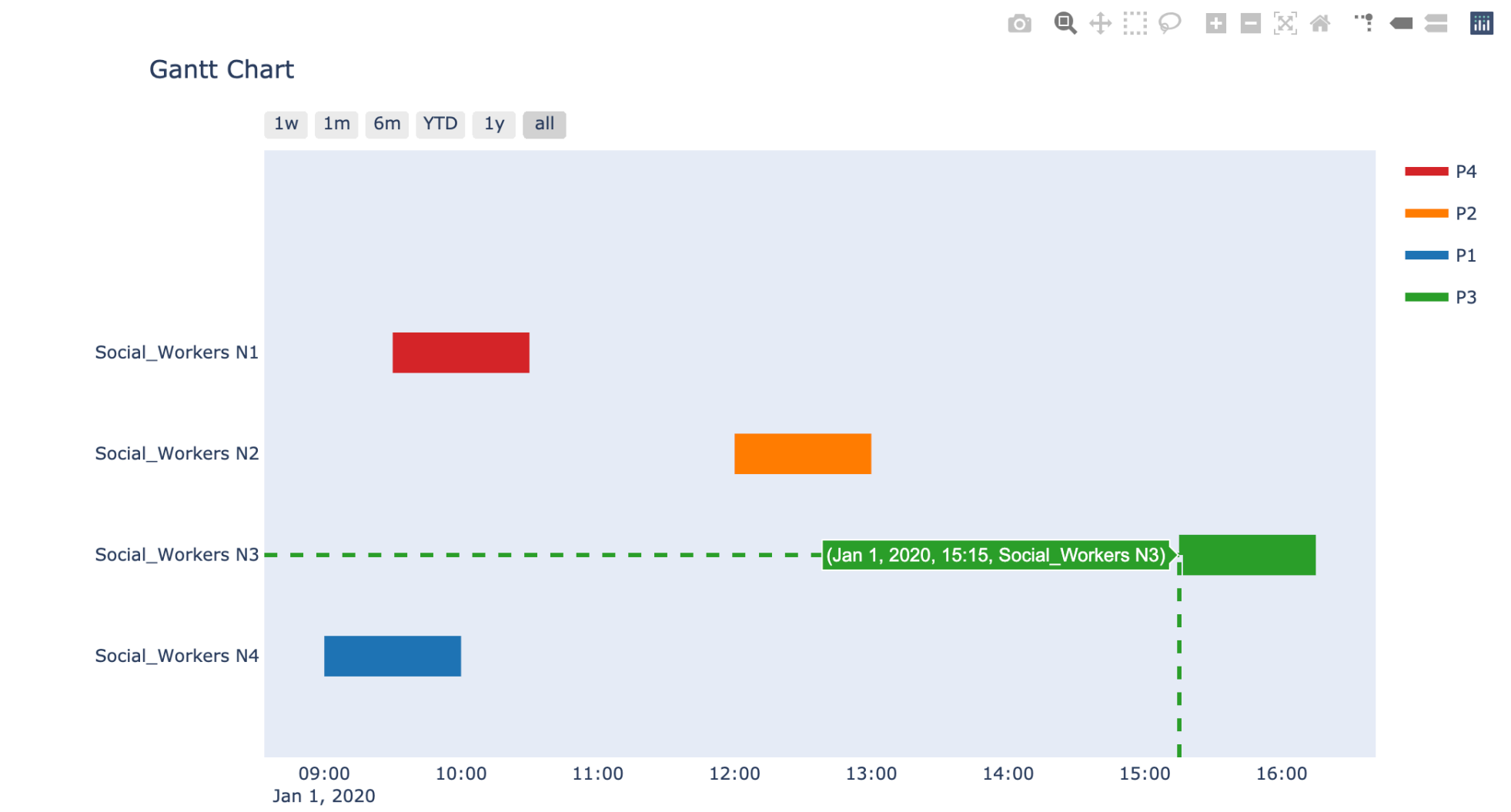

**Fig. 18.** Result of the social workers' problem algorithm using the *ibmq_16_melbourne*. In this scenario, it is showed the hypothesis of 4 patients and four social workers. In this case, it is verified that the implementation and restrictions work very well.

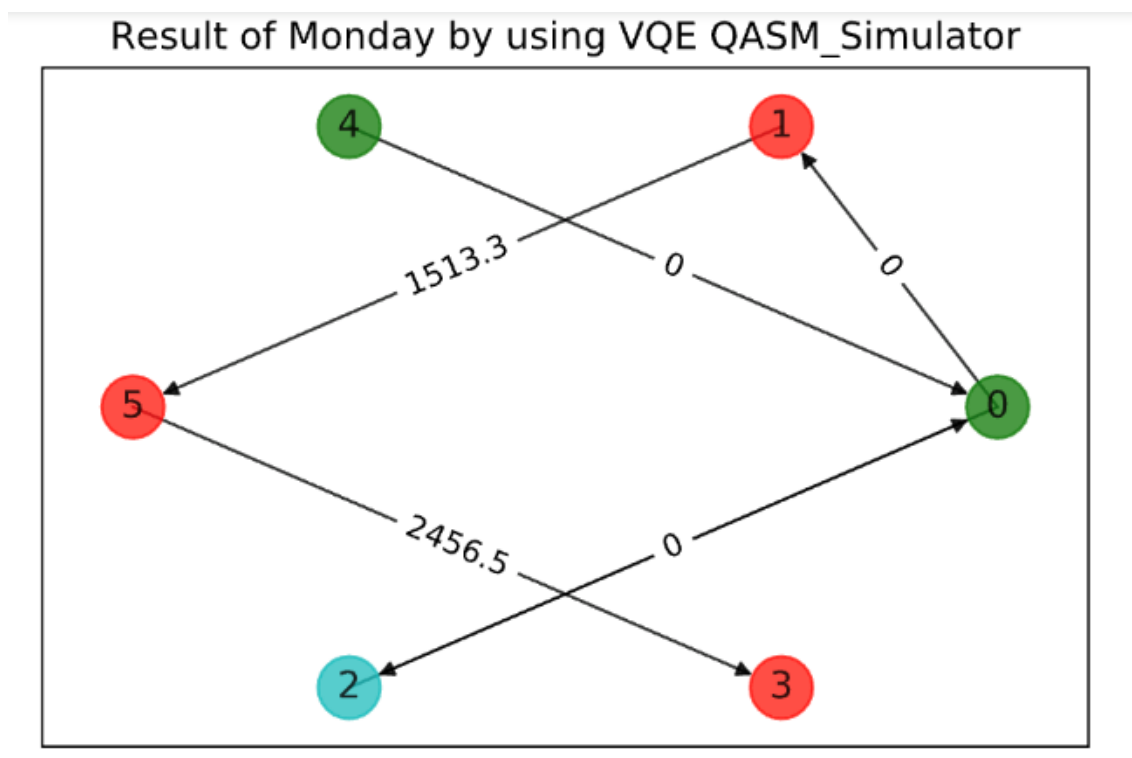

```
Result of Monday's path
energy: -427701.86611328117
time: 1930.61497092247
solution objective: 125849.38388672081
solution: [1 0 0 1 0 0 1 0 0 1 0 0 0 0 1 1 0 0]
Social Worker TimeTable
-----------------------------------
Social Worker 1:
        Patient 1: 09:00-10:00
        Patient 5: 11:45-12:45
        Patient 3: 15:15-16:15
-----------------------------------
Social Worker 2:
        Patient 2: 12:00-13:00
```

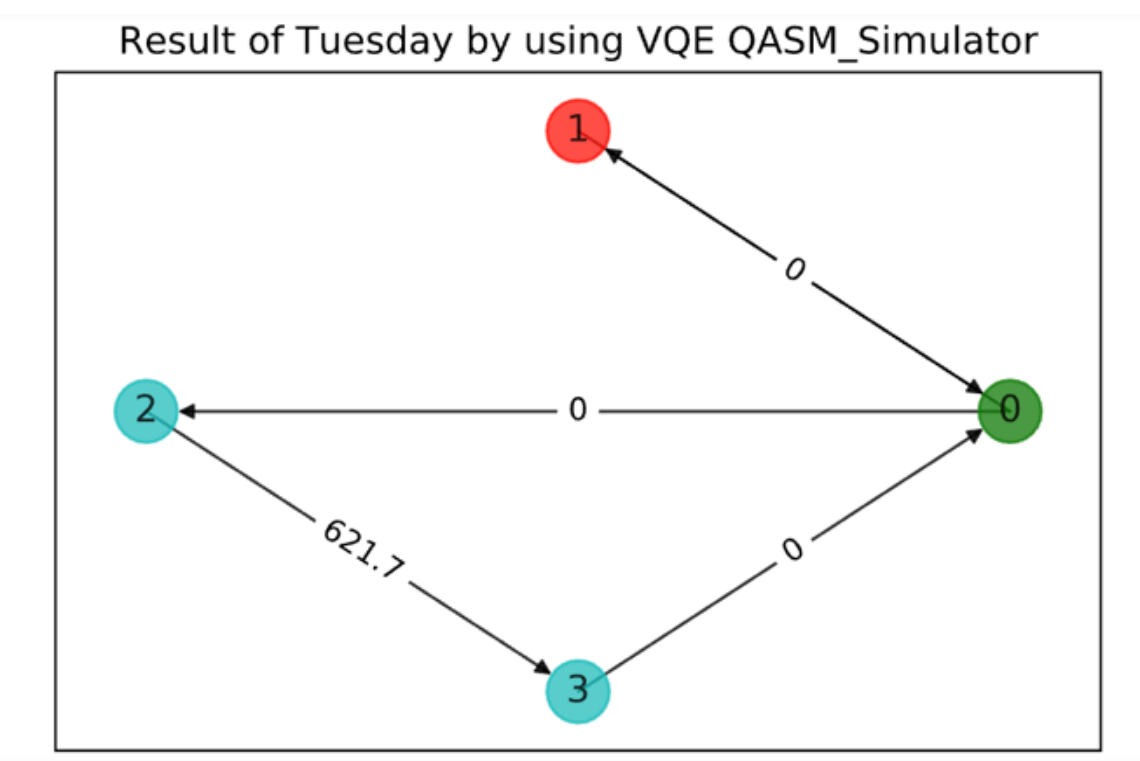

```
Result of Tuesday's path
energy: -118800.83974609376
time: 78.25093412399292
solution objective: 38412.960253906116
solution: [1 1 0 1 0 0 0 1 1]
Social Worker TimeTable
-----------------------------------
Social Worker 1:
          Patient 1: 09:00-10:00
-----------------------------------
Social Worker 2:
          Patient 2: 14:45-15:45
          Patient 3: 16:30-17:30
```

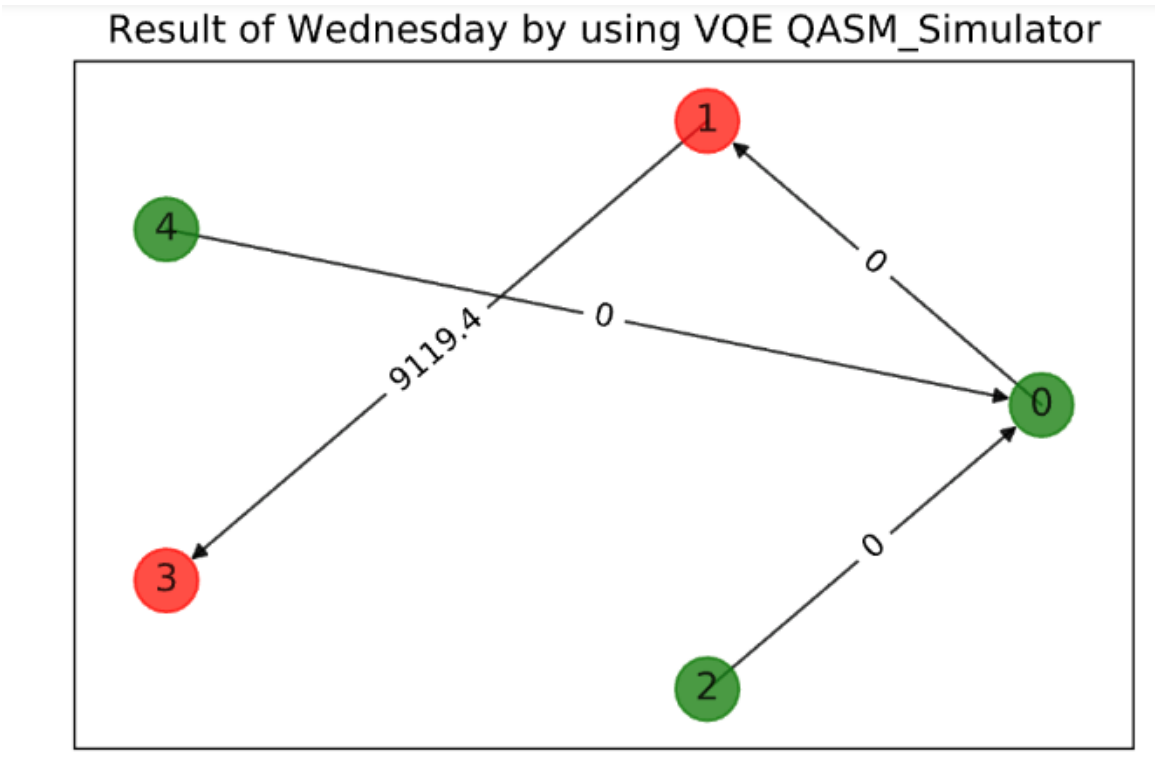

```
Result of Wednesday's path
energy: -252857.77705078124
time: 394.21149587631226
solution objective: 177953.72294921876
solution: [1 0 0 0 0 0 1 0 1 0 0 0 1]
Social Worker TimeTable
----------------------------------
Social Worker 1:
          Patient 1: 09:00-10:00
          Patient 3: 15:45-16:45
```

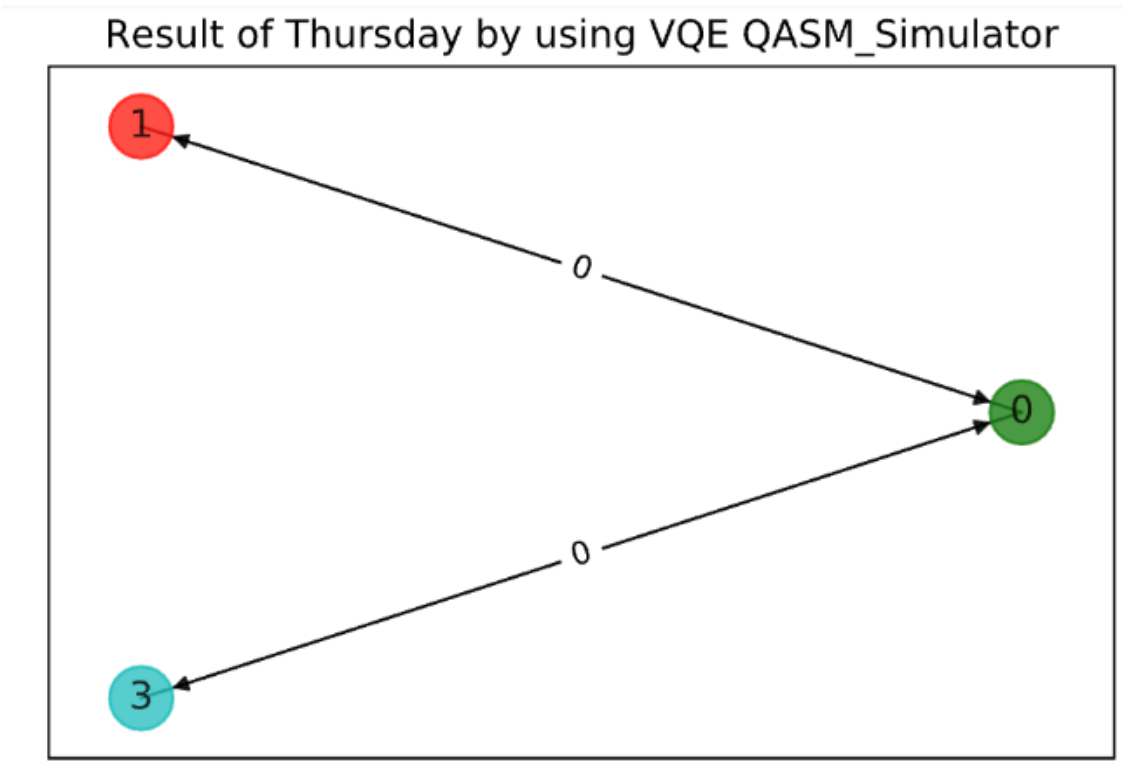

```
Result of Thursday's path
energy: -32104.224511718738
time: 15.917151927947998
solution objective: 14746.625488281225
solution: [1 1 1 0 1]
Social Worker TimeTable
----------------------------------
Social Worker 1:
          Patient 1: 09:00-10:00
----------------------------------
Social Worker 2:
          Patient 3: 15:00-16:00
```

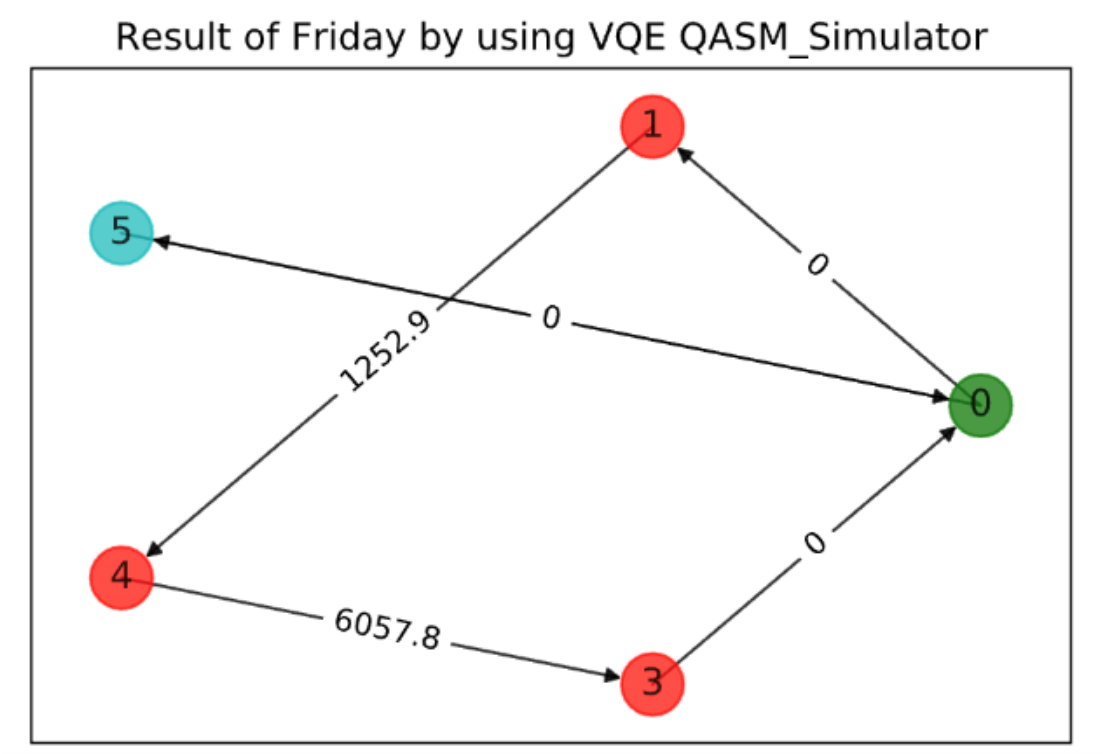

```
energy: -281565.03271484375
time: 445.0221631526947
solution objective: 88684.11728515604
solution: [1 1 0 0 0 0 1 0 1 0 0 1 1]
Social Worker TimeTable
----------------------------------
Social Worker 1:
          Patient 1: 09:00-10:00
          Patient 4: 11:30-12:30
          Patient 3: 17:00-18:00
----------------------------------
Social Worker 2:
          Patient 5: 11:30-12:30
```

**Fig. 19.** Result of the quantum simulator with the backend *qasm_simulator* of the Social Workers' scheduler. This is the output of the algorithm through the quantum execution line without going through the system based on the cases already previously calculated. This result is from the first experiments before having some memories of cases already solved in the solution space.

Result of our QSW algorithm based on CBR. Total_weight: 2765.5

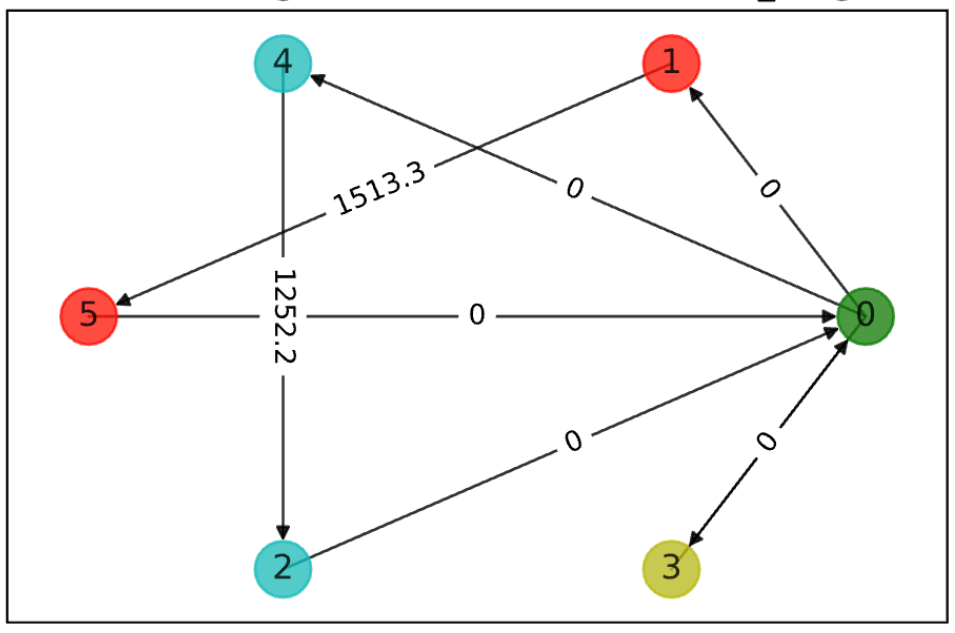

```
Result of Monday's path
Social Worker TimeTable
-----------------------------------
Social Worker 1:
          Patient 1: 09:00-10:00
          Patient 5: 11:45-12:45
Social Worker 2:
          Patient 4: 09:30-10:30
          Patient 2: 12:00-13:00
Social Worker 3:
          Patient 3: 15:15-16:15
```

Result of our QSW algorithm based on CBR. Total_weight: 0

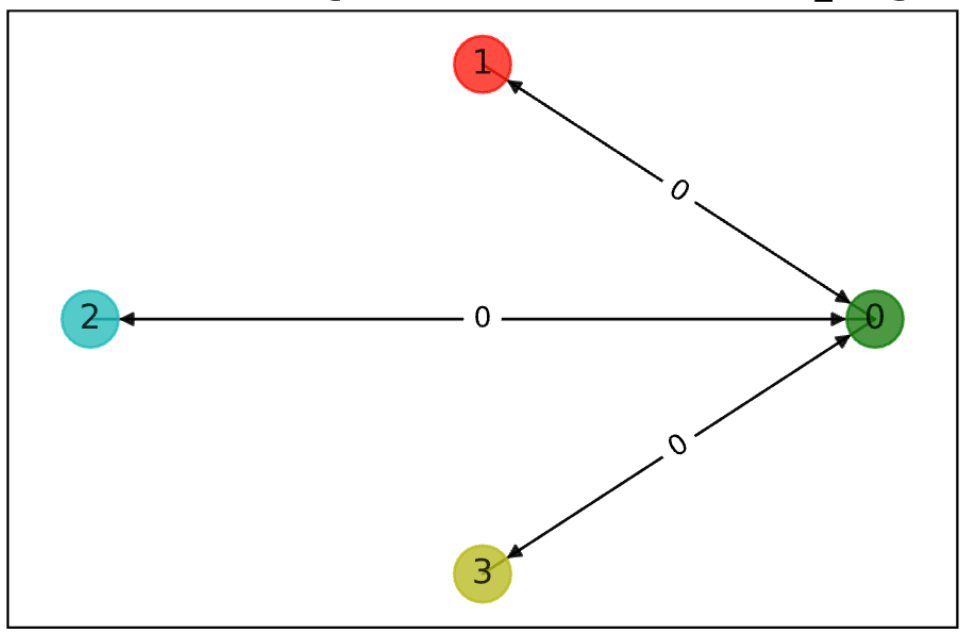

```
Result of Tuesday's path
Social Worker TimeTable
-----------------------------------
Social Worker 1:
          Patient 1: 09:00-10:00
Social Worker 2:
          Patient 2: 14:45-15:45
Social Worker 3:
          Patient 3: 16:30-17:30
```

Result of our QSW algorithm based on CBR. Total_weight: 315.4

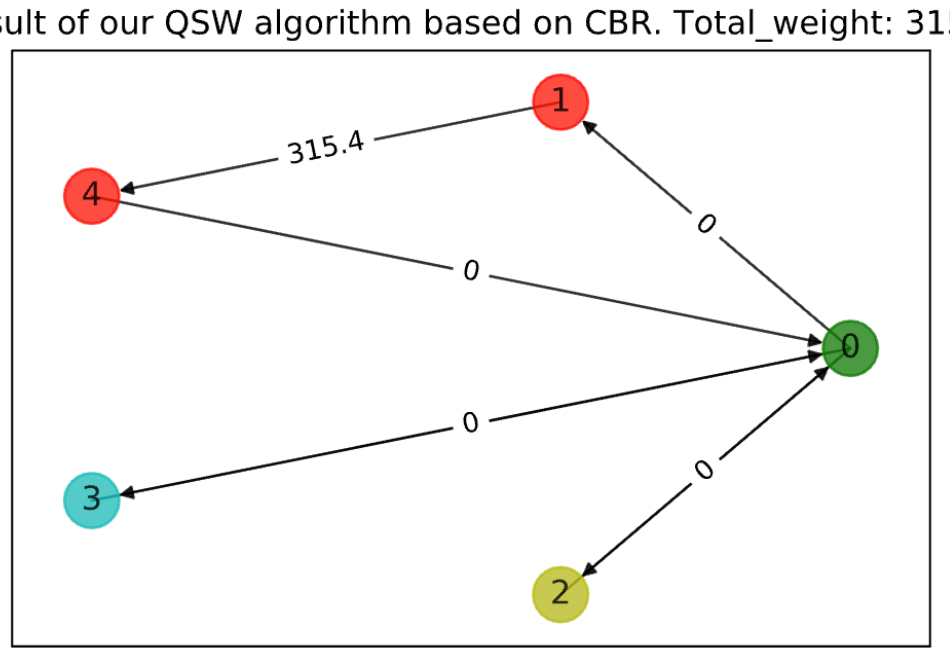

```
Result of Wednesday's path
Social Worker TimeTable
-----------------------------------
Social Worker 1:
          Patient 1: 09:00-10:00
          Patient 4: 10:15-11:15
Social Worker 2:
          Patient 3: 15:45-16:45
Social Worker 3:
          Patient 2: 16:00-17:00
```

Result of our QSW algorithm based on CBR. Total_weight: 0

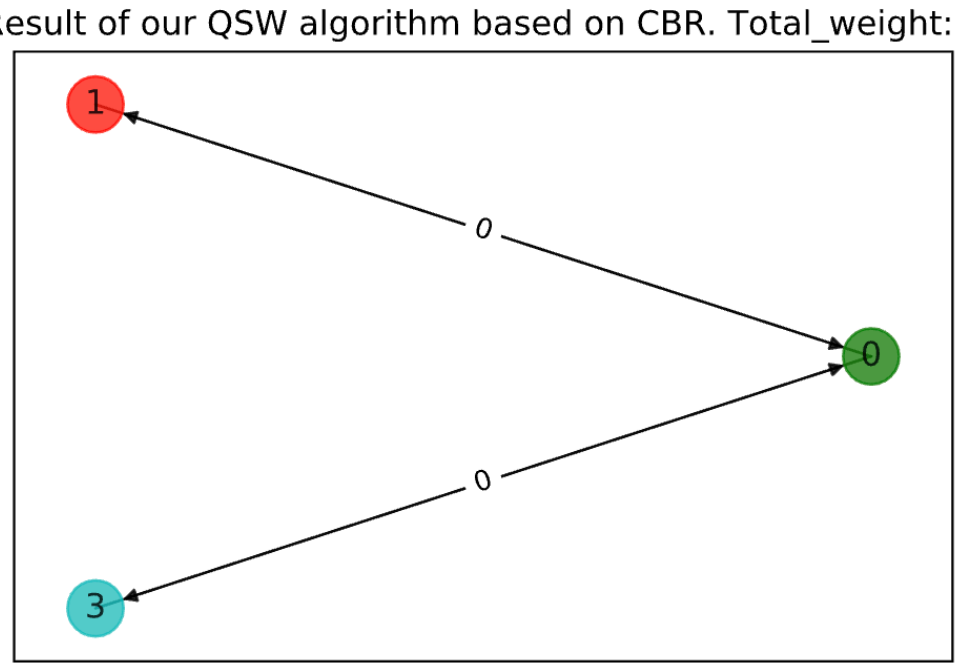

```
Result of Thursday's path
Social Worker TimeTable
-----------------------------------
Social Worker 1:
          Patient 1: 09:00-10:00
Social Worker 2:
          Patient 3: 15:00-16:00
```

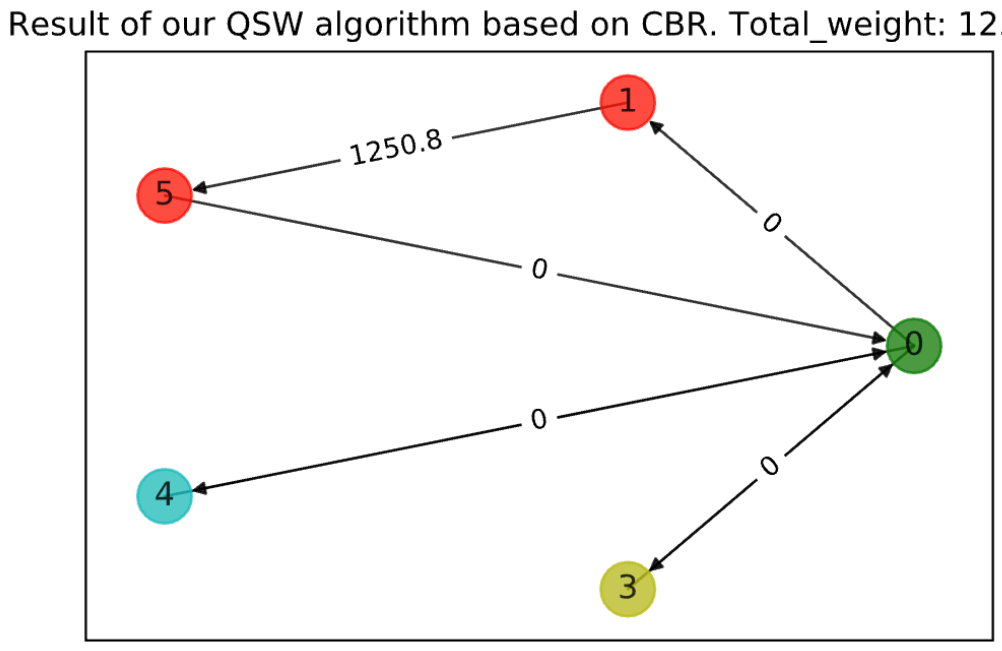

```
Result of Friday's path
Social Worker TimeTable
----------------------------------
Social Worker 1:
        Patient 1: 09:00-10:00
        Patient 5: 11:30-12:30
Social Worker 2:
        Patient 4: 11:30-12:30
Social Worker 3:
        Patient 3: 17:00-18:00
```

**Fig. 20.** Results of our quantum algorithm based on the recovery of the past schedule of the social workers with the weighting factor. In this case, the optimal schedules of 3 social workers related to 5 patients are sought in a random location. It is worth mentioning that before executing this case, we had already achieved 128 previous cases. Thus, the case-based reasoning system already has a memory (although small) of cases. For this, the changes introduced in this experiment are not so drastic.

**Table 3.** Comparative Table of five scenarios taking into account the number of patients and the number of qubits, with which the number of layers will be calculated. Depending on the type of 'linear' entanglement, the number of times it solves the problem of social workers recovering the solution based on cases already solved without optimizing the case again, and it is reoptimized or calculated from zero (Top-Down) the project.

| # Patients | # Social Workers | # Qubits | # Layers | Linear entanglement | # Total cases | Success rate recovering the solution without optimizing. | Success rate recovering the solution and optimizing from the last optimal point. | The success rate from "reshaping" the entire problem from the scratch |
|---|---|---|---|---|---|---|---|---|
| 3 | 2 | 6 | 2 | YES | 243 | 58,53% | 31,67% | 9,8% |
| 4 | 3 | 12 | 3 | YES | 243 | 56,60% | 34,63% | 8,8% |
| 5 | 2 | 20 | 4 | YES | 243 | 53,53% | 30,32% | 16,2% |
| 5 | 3 | 20 | 5 | YES | 243 | 48,17% | 33,28% | 18,6% |
| 5 | 4 | 20 | 10 | YES | 243 | 49,91% | 39,24% | 10,9% |

**Table 4.** Comparative Table of five scenarios taking into account the number of patients and the number of qubits, with which the number of layers will be calculated. Depending on the type of 'full' entanglement, the number of times it solves the problem of social workers recovering the solution based on cases already solved without optimizing the case again, and it is reoptimized or calculated from zero (Top-Down) the project.

| # Patients | # Social Workers | # Qubits | # Layers | Full entanglement | # Total cases | Success rate recovering the solution without optimizing. | Success rate recovering the solution and optimizing from the last optimal point. | The success rate from "reshaping" the entire problem from the scratch |
|---|---|---|---|---|---|---|---|---|
| 3 | 2 | 6 | 2 | YES | 243 | 61,13% | 25,87% | 13,0% |
| 4 | 3 | 12 | 3 | YES | 243 | 59,20% | 28,37% | 12,4% |
| 5 | 2 | 20 | 4 | YES | 243 | 58,70% | 23,97% | 17,3% |
| 5 | 3 | 20 | 5 | YES | 243 | 59,38% | 27,15% | 13,5% |
| 5 | 4 | 20 | 10 | YES | 243 | 59,61% | 32,79% | 7,6% |

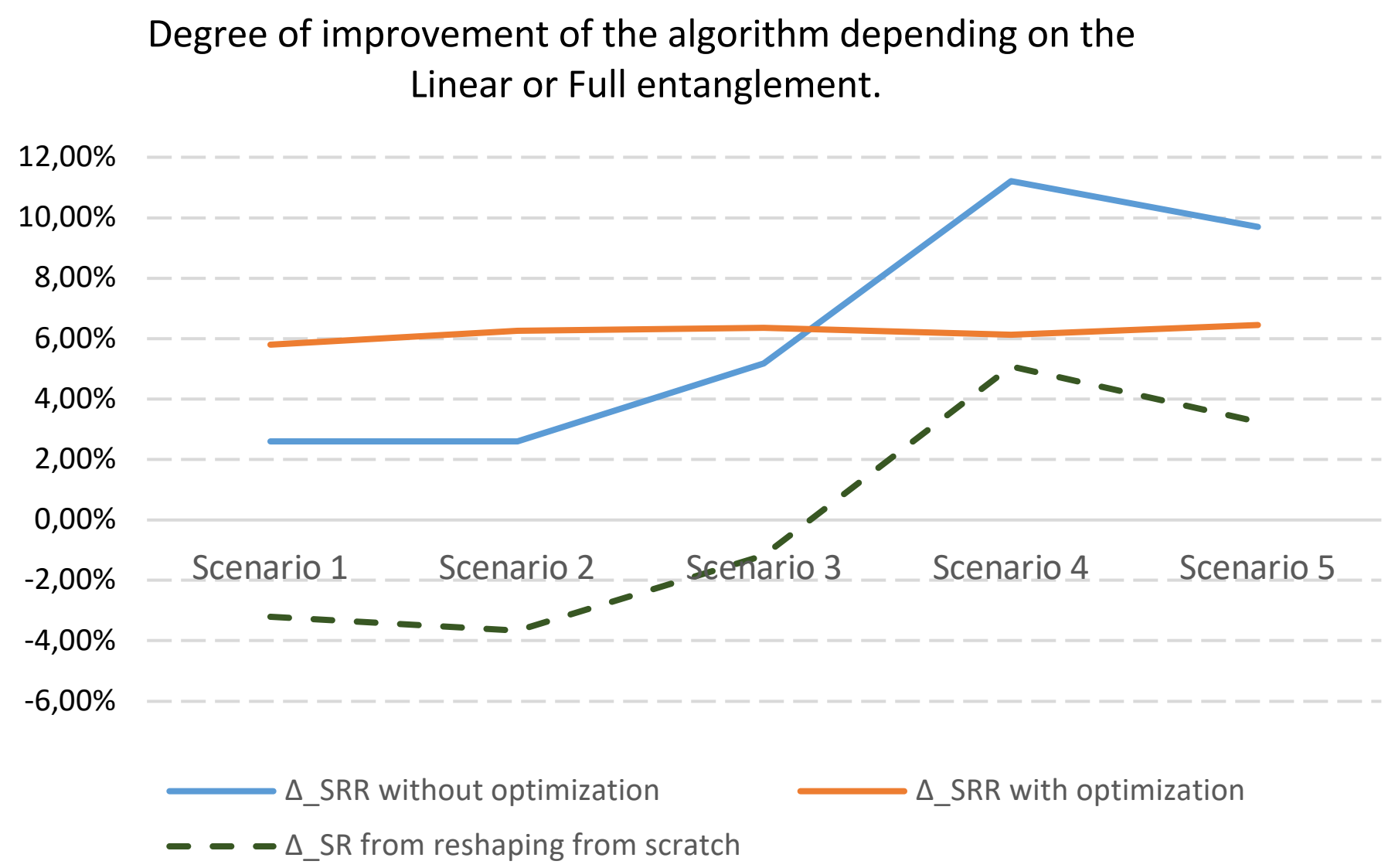

**Fig. 21.** Comparative graphs in five scenarios that analyze the improvements of the three techniques based on linear or full entanglement. Through the graphs, one observes the improvement of the algorithm based on the recovery of cases without the need to optimize the recovered case. This improvement is around 6%. On the other hand, the branding of the code based on reshaping is kept in absolute value. It is compensated with the variation of the whiteness of the optimization of the last optimum point.

## 10 Discussion

One of the objectives of this article is to achieve an adequate computational methodology that allows, given a specific problem, to be able to solve it in a general way to achieve a system both efficient at the level of quantum resources (reducing costs) and as intelligent. Since this system will end up learning by analogy, and it will make it easier for us not to have to recalculate (Top-Down), the new entries from scratch.

To find this new methodology, about 20 ansatzes were studied and, after the results obtained, the one proposed in figure (6) was chosen because it shows that the VQE and its variation performed work satisfactorily and enhance the QML era.
The social workers' problem is a fairly generic type of problem since it combines at least two kinds of issues, routing and planning. Besides, considering the approach made and taking into account the new restrictions and conditions, A hybrid approach has been proposed that can become very powerful for solving NP-Hard combinatorial problems, supervised, unsupervised, and case-based systems.
The problem of generating optimal schedules for social workers who visit their patients at home can be solved in a specific way by leveraging a QUBO formulation, improving as it has been done in other articles or solving it with QAOA or other techniques variational (Grover [26], Admm [27], etc.). With QAOA, the PQC as particulate with the information on the problem, see in figure (5). Remember that when using VQE, it is not necessary to customize the PQC. The best PQC is chosen that approximately guarantees us the best possible state vector in Hilbert space, and with this, the variational algorithm is called. This is the worth of using VQE or its derivatives. That is, the VQE does not influence the PQC, nor does it depend directly on our problem. But, on the contrary, the QAOA does depend on the problem to be solved, if we seek efficiency.
But solving the problem in that way (only Top-Down view), and in case of drastic changes in the approach, we should change part of our algorithm or in the best of circumstances, require a computational cost in each associated execution to the NP-Hard problem. The approach that it is proposed in this article allows us to guarantee at least two characteristics. First, it has been verified that it was exclusively used as the resource necessary for resolution. This allows us to make efficient use of the computational resource, and second, above all, it lays the basis for generalizing problem-solving based on routing and planning. In conclusion, it has been presented some useful data to analyze the efficiency of the contribution, recalling the results obtained on the generalization of (Fig. 11.).
Tables (3) and (4) together with figure (21) summarize the results of the quantum algorithm based on case reasoning. Everything that was developed in this article forms a solid base towards Quantum Cases-Based Reasoning (QCBR), and the results obtained encourage us to create hybrid and intelligent systems.

Since in the case for $n = 5$, the number of qubits necessary is $n(n-1) = 20\ qubits$. Not many more tests could be done for values of $n$ greater than 5, since the computer on which the algorithm was tested takes too long.
With this work, it is sought to offer cities an instrument [28] which could optimize the costs allied to the management of social workers and improve social gains. This work could also be a starting point for many countries in Africa that are seizing the opportunity of the mobile technology revolution [29] to develop and increase their industrious and e-health system [30] [31].
One would like to add that suggested formulation (12, 13) is not only specific to the proposed problem. It can be used to solve any family planning, scheduling and routing problem related to a list of tasks, restrictions, allocation of resources on location and time. The test performed and showed in the figure (22) allows to analyze the behaviour of the formulation with the variation of the correction factor $\varepsilon$. One can observe how the time window $T_{ij} = (\tau_{i-}\tau_j)$ adapts perfectly at the extremes to the cost variable in the distance. This achievement is due to the chosen quadratic function (13). One wants it to be adapted in this way so that our resultant function (12) weights together the short distances and time as well as the long distances and late times.

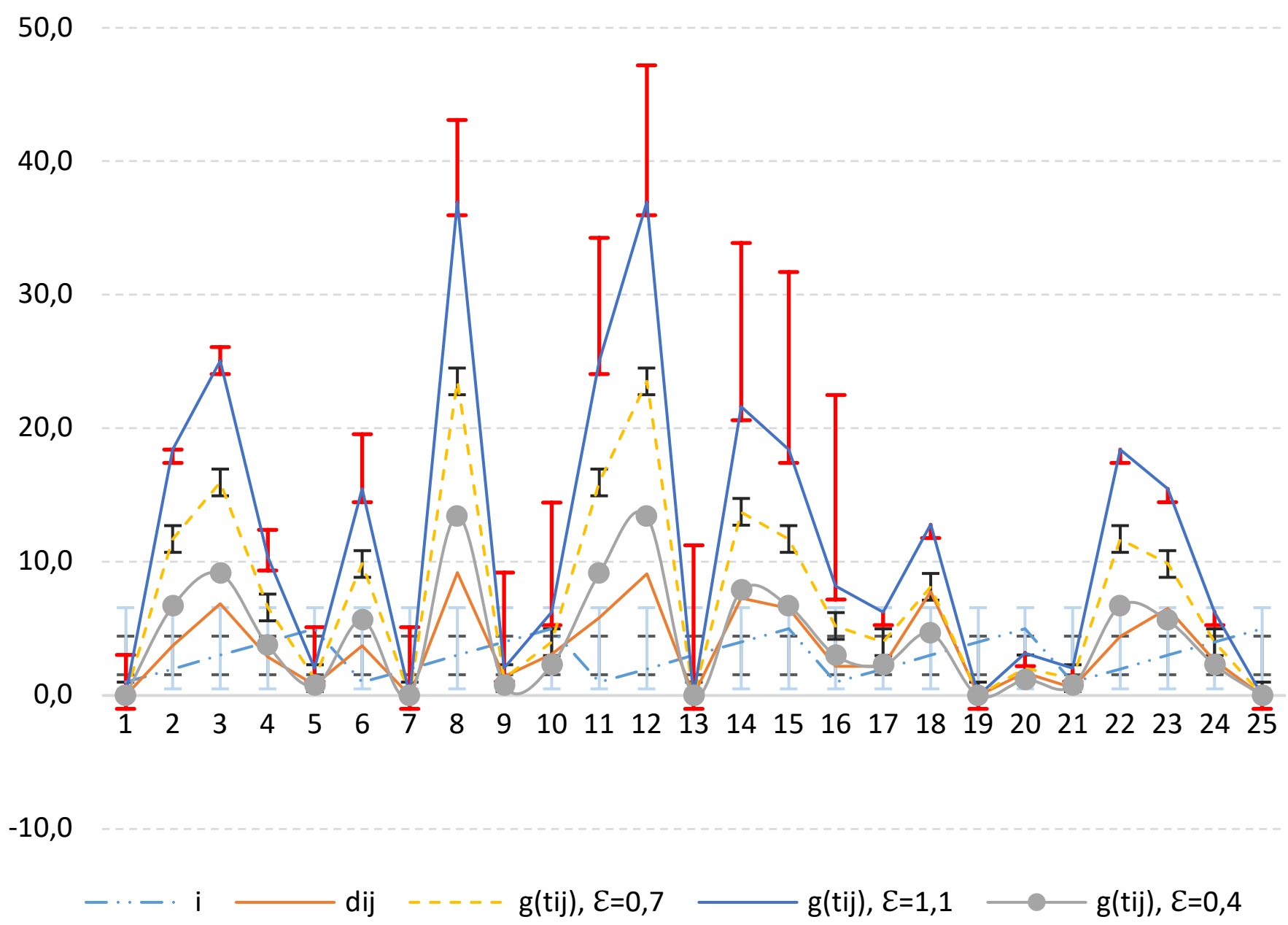

**Fig. 22.** Comparison of performance with standard deviation error bar on the three mappings of our proposed formulation (12,13). The Standard Deviation expected total anneal time for 98% percent success for each mapping value, with the best $\varepsilon$ for each shoot are shown. Our optimal case is for $\varepsilon = 0{,}7$. The most representative cases are for $\varepsilon = 0{,}4$, $\varepsilon = 0{,}7$ and $\varepsilon = 1{,}1$

Other functions can be studied to have a test bench with which to compare the final results.

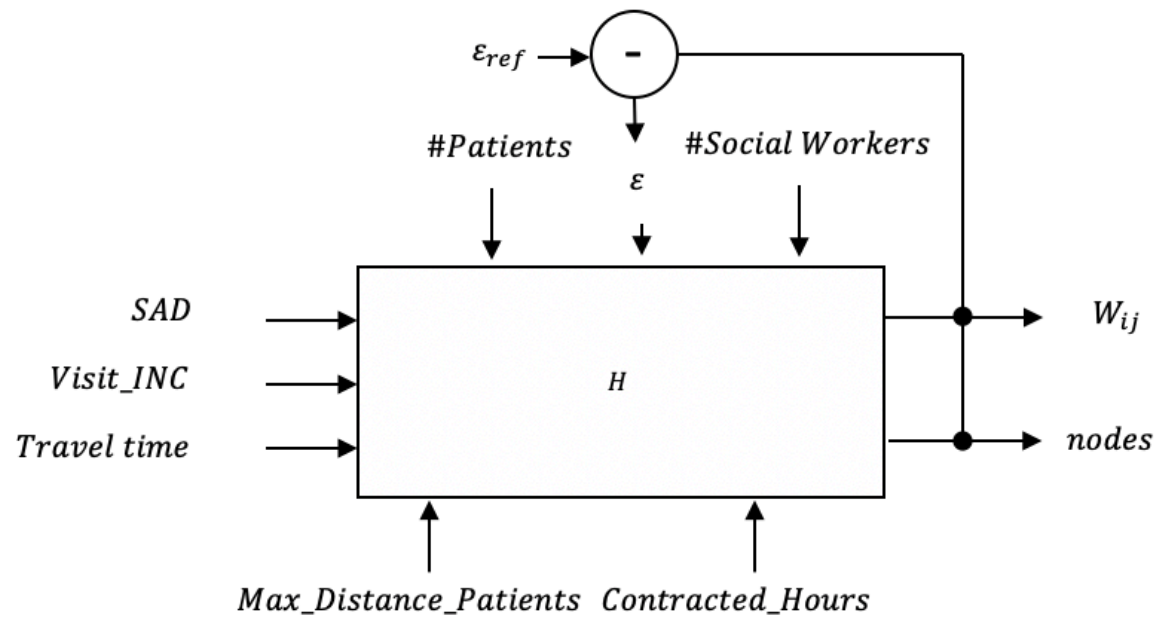

**Fig. 23.** Block diagram of the control system that would allow better control over $\varepsilon$.

Another improvement that has been introduced based on previous works is to design a control system that improves the weighting parameter of the time window function (see figure (23)).

## 11 Conclusion and further works

A parameterized quantum circuit has been proposed in conjunction with Variational-Quantum-Eigensolver (VQE) combined with a case-based system to create a smart approach to generate optimal social worker schedules.

The structure of the system is as follows: the data is loaded into the first gate $U_3$ and the processing parameters $(x;\theta)$ are loaded into the remaining quantum gates several times (max_iteration) along the circuit by rotations of the 15 qubits in the first gate. The processing parameters of these rotations are different in each load, and we optimize them using a classic minimization algorithm (COBYLA and SPSA).
A similarity function has been defined that explores both the declaration space and the solution space by establishing and implementing a case-based system that helps make the overall system more efficient. Although it is a base, the system works very well, since our QCBR solves 89% of the 243 cases, and the remaining 11% go through the standard way (Top-Down) of the Hamiltonian calculation. It is also of interest to remember that the computational cost of the resulting system is reduced by 49% concerning what it would be if the system had not been implemented. With this and with the era (NISQ) we are in, this hybrid system, in addition to being functional, helps us reduce the cost of shared quantum resources.

In short, it can be said that the proposed and implemented a hybrid quantum-classical system helps to generate optimal schedules for the problem of social workers who visit their patients at home.

As a future direction, one of the tasks that it will be intended to carry out is to design a quantum machine learning system to optimize the exploration of the statement and solution spaces for the QCBR, to improve access time in the memory. In addition to doing a comparative study with several classical techniques. Taking full advantage of the innate characteristics of the quantum computer, which is to perform operations in the Hilbert vector space (inner and outer product) in a simple (native) way; tremendously complicated things for classic computers and that are the foundations of A.I. techniques (kernel, feature map, etc.).

**Acknowledgements**

The authors greatly thank the IBMQ team, mainly Steve Wood. P.A. thanks Jennifer Ramírez Molino and Adrian Perez-Salinas for their support and comments on the manuscript.